\documentclass{emulateapj}

\usepackage{natbib}
\usepackage{graphicx}
\usepackage{latexsym,amssymb}
\usepackage{amsmath}

\newcommand{\kms}{km\,s$^{-1}$}
\newcommand{\dgr}{$^\circ$}
\newcommand{\Msun}{M$_\odot$}

\newcommand{\Lsunpcsq}{L$_\odot$\,pc$^{-2}$}
\newcommand{\Msunpcsq}{M$_\odot$\,pc$^{-2}$}

\newcommand{\MLsunI}{$\Upsilon_{\odot,I}$}
\newcommand{\kmsM}{km\,s$^{-1}$\,Mpc$^{-1}$}      
\newcommand{\Ls}{L$\star$}      

\newcommand{\Reff}{\ensuremath{R_e}}
\newcommand{\Rein}{\ensuremath{R_E}}
\newcommand{\Mein}{\ensuremath{M_E}}
\newcommand{\Sigc}{\ensuremath{\Sigma_c}}
\newcommand{\Rc}{\ensuremath{R_c}}
\newcommand{\Mc}{\ensuremath{M_c}}

\newcommand{\MLein}{\ensuremath{\Upsilon_E}}
\newcommand{\MLdyn}{\ensuremath{\Upsilon_\mathrm{dyn}}}
\newcommand{\MLstar}{\ensuremath{\Upsilon_\star}}
\newcommand{\MLtot}{\ensuremath{\Upsilon_\mathrm{tot}}}
\newcommand{\MLj}{\ensuremath{\Upsilon_j}}
\newcommand{\Mbh}{\ensuremath{M_\mathrm{BH}}}
\newcommand{\du}{\mathrm{d}}               
\newcommand{\HI}{\ion{H}{1}}
\newcommand{\gmos}{\texttt{GMOS}}          
\newcommand{\castles}{\texttt{CASTLES}}    
\newcommand{\hst}{HST}                     
\newcommand{\wmap}{WMAP}                   
\newcommand{\gemini}{GEMINI}               
\newcommand{\feh}{\mathrm{[Fe/H]}}         

\shorttitle{Stellar dynamics and gravitational lensing}
\shortauthors{van de Ven et al.}

\begin{document}

\title{The Einstein Cross: constraint on dark matter from stellar
  dynamics and gravitational lensing}

\author{
Glenn van de Ven\altaffilmark{1,2},
Jes\'us Falc\'on--Barroso\altaffilmark{3,4},
Richard M.\ McDermid\altaffilmark{5},
Michele Cappellari\altaffilmark{6},
Bryan W.\ Miller\altaffilmark{7},
P.\ Tim de Zeeuw\altaffilmark{8,9}
}

\altaffiltext{1}{Max Planck Institute for Astronomy, K\"onigstuhl 17,
  69117 Heidelberg, Germany: glenn@mpia.de}

\altaffiltext{2}{Institute for Advanced Study, Einstein Drive,
  Princeton, NJ 08540, USA; Hubble Fellow}

\altaffiltext{3}{Instituto de Astrof\'isica de Canarias. V\'ia
  L\'actea s/n, La Laguna. Tenerife. Spain}

\altaffiltext{4}{European Space Agency / ESTEC, Keplerlaan 1, 2200 AG
  Noordwijk, The Netherlands}

\altaffiltext{5}{Gemini Observatory, 670 N.\ A'ohoku Place Hilo,
  Hawaii, 96720, USA}

\altaffiltext{6}{Sub-Department of Astrophysics, University of Oxford,
  Denys Wilkinson Building, Keble Road, Oxford OX1 3RH, United
  Kingdom}

\altaffiltext{7}{Gemini Observatory, Casilla 603, La Serena, Chile}

\altaffiltext{8}{European Southern Observatory, Karl-Schwarzschild
  Strasse 2, 85748 Garching bei M\"unchen, Germany}

\altaffiltext{9}{Sterrewacht Leiden, Leiden University, Niels Bohrweg
  2, 2333 CA Leiden, The Netherlands}

\begin{abstract}
  We present two-dimensional line-of-sight stellar kinematics of the
  lens galaxy in the Einstein Cross, obtained with the \gemini\ 8m
  telescope, using the \gmos\ integral-field spectrograph. The stellar
  kinematics extent to a radius of $4$\arcsec\ (with $0.2$\arcsec\
  spaxels), covering about two-thirds of the effective (or half-light)
  radius $\Reff \simeq 6$\arcsec\ of this early-type spiral galaxy at
  redshift $z_l \simeq 0.04$, of which the bulge is lensing a
  background quasar at redshift $z_s \simeq 1.7$.  The velocity map
  shows regular rotation up to $\sim 100$\,\kms\ around the minor axis
  of the bulge, consistent with axisymmetry. The velocity dispersion
  map shows a weak gradient increasing towards a central
  ($R<1$\arcsec) value of $\sigma_0 = 170 \pm 9$\,\kms.
  We deproject the observed surface brightness from \hst\ imaging to
  obtain a realistic luminosity density of the lens galaxy, which in
  turn is used to build axisymmetric dynamical models that fit the
  observed kinematic maps. We also construct a gravitational lens
  model that accurately fits the positions and relative fluxes of the
  four quasar images. We combine these independent constraints from
  stellar dynamics and gravitational lensing to study the total mass
  distribution in the inner parts of the lens galaxy. \looseness=-1
  
  We find that the resulting luminous and total mass distribution are
  nearly identical around the Einstein radius $\Rein = 0.89$\arcsec,
  with a slope that is close to isothermal, but which becomes
  shallower towards the center if indeed mass follows light. The
  dynamical model fits to the observed kinematic maps result in a
  total mass-to-light ratio $\MLdyn = 3.7 \pm 0.5$\,\MLsunI\ (in the
  $I$-band). This is consistent with the Einstein mass $\Mein = 1.54
  \times 10^{10}$\,\Msun\ divided by the (projected) luminosity within
  $\Rein$, which yields a total mass-to-light ratio of $\MLein =
  3.4$\,\MLsunI, with an error of at most a few per cent. We estimate
  from stellar populations model fits to colors of the lens galaxy a
  stellar mass-to-light ratio $\MLstar$ from $2.8$ to $4.1$\,\MLsunI.
  Although a \emph{constant} dark matter fraction of 20 per cent is
  not excluded, dark matter may play no significant role in the bulge
  of this $\sim$\Ls\ early-type spiral galaxy. 
\end{abstract}

\keywords{gravitational lensing --- stellar dynamics --- galaxies:
  photometry --- galaxies: kinematics and dynamics --- galaxies:
  structure}

\section{Introduction}
\label{sec:intro}

In the cold dark matter (CDM) paradigm for galaxy formation
\citep[e.g.][]{2002SciAm.286..36K}, galaxies are embedded in extended
dark matter distributions with a specific and (nearly) universal
radial profile \citep[e.g.][]{1997ApJ...490..493N,
  2005ApJ...624L..85M}. Measurements of rotation curves from neutral
hydrogen (\HI) observations in the outer parts of late-type galaxies
have provided evidence for the presence of dark matter in these
systems already more than two decades ago
\citep[e.g.][]{1985ApJ...295..305V}. In the outer parts of early-type
galaxies, however, cold gas is scarce \citep[but see
e.g.][]{1994ApJ...436..642F, 1997AJ....113..937M,
  2008MNRAS.383.1343W}, so evidence for a dark matter halo has to come
from other tracers, such as kinematics of stars, planetary nebulae and
globular clusters \citep[e.g.][]{1995ApJ...441L..25C,
  2001AJ....121.1936G, 2003Sci...301.1696R, 2003ApJ...591..850C} or
hot X-ray gas \citep[e.g.][]{2006ApJ...636..698F,
  2010MNRAS.403.2143H}.  However, these tracers are not always
(sufficiently) available, the observations are often challenging, and
the interpretation modeling dependent. Also in the inner parts of
galaxies the amount, shape and profile of dark matter is still poorly
known \citep[e.g.][]{2004IAUS..220...53P}, even though stellar
kinematics are readily available.

A fundamental problem in using stellar kinematics (as well as other
collisionless kinematic tracers) for this purpose is the
mass-anisotropy degeneracy: a change in the measured line-of-sight
velocity dispersion can be due to a change in total mass, but also due
to a change in velocity anisotropy. Both effects can be disentangled
by measuring also the higher-order velocity moments
\citep{1987MNRAS.224...13D, 1993ApJ...407..525V, 1993MNRAS.265..213G},
but only the inner parts of nearby galaxies are bright enough to
obtain the required high-quality kinematic measurements
\citep[e.g.][]{1991MNRAS.253..710V, 2001AJ....121.1936G,
  2006MNRAS.366.1126C}. A unique alternative to break the
mass-anisotropy degeneracy is provided by gravitational lensing. In
case of strong lensing, the mass of a foreground galaxy bends the
light of a distant bright object behind it, resulting in multiple
images. From the separation and fluxes of the images the total mass
distribution of the lens galaxy can be inferred directly. The velocity
anisotropy then can be determined from the observed velocity
dispersion, without the need for higher-order velocity moments.

\citeauthor{2004ApJ...611..739T} (\citeyear{2004ApJ...611..739T}, and
references therein) have applied this approach to several strong
lensing systems, of which 0047-281 \citep{2003ApJ...583..606K} is the
best constrained case, with three radially separated velocity
dispersion measurements extending to about the effective radius of the
lens galaxy. They measure the mass within the Einstein radius by
fitting a singular isothermal ellipsoid to the positions of the quasar
images.  This Einstein mass is used to set the amplitude of the total
(stellar and dark) matter distribution, which they assume to be
spherical. The constant stellar mass-to-light ratio $\MLstar$
determines the contribution of the stars with a fixed radial profile,
with the remainder due to dark matter with a single power-law profile
with slope $\gamma$. They then compare the dispersion profile
predicted by the spherical Jeans equations, for an ad-hoc assumption
of the velocity anisotropy $\beta$, with the observed dispersion
measurements. Based on a reasonably constrained $\MLstar$ and an upper
limit on $\gamma$, they conclude that a significant amount of dark
matter is present in the inner parts of the lens galaxy, with a slope
shallower than the nearly isothermal total mass distribution. An
additional (external) constraint on $\MLstar$ (or $\gamma$) is needed
to go beyond this limit on the dark matter distribution. Even so,
their results are limited by too few kinematic constraints (which
leaves the anisotropy degenerate), and by the use of a simple
spherical dynamical model.

Clearly, most lens galaxies are significantly flattened and so cannot
be well-described by spherical models. Non-spherical models provide a
more realistic description of the lens galaxy, but the increase in
freedom requires also (significantly) more spatially resolved
kinematic measurements to constrain them. Only a few of the known
strong gravitational lens systems are close enough to obtain such
kinematic data. Even then, one can make (ad-hoc) assumptions on the
velocity distribution, such as velocity isotropy in the meridional
plane of an axisymmetric model. The latter restriction to a
distribution function of two (instead of three) integrals of motion
allows for the recovery of the intrinsic shape and mass distribution
in the presence of two-dimensional kinematic data
\citep{2007ApJ...666..726B, 2008MNRAS.384..987C, 2009MNRAS.399...21B},
possibly even without the additional constraint provided by
gravitational lensing \citep[e.g.][]{2006MNRAS.366.1126C}.

In this paper, we relax the two-integral assumption and find that one
can still derive independent determinations of the intrinsic mass
distribution from both gravitational lensing and stellar dynamics with
two-dimensional kinematic data. We have observed the gravitational
lens system QSO\,2237+0305, well-known as the Einstein Cross, with the
integral-field spectrograph \gmos\ on the \gemini-North Telescope.  We
fit the extracted stellar velocity and velocity dispersion maps of the
inner parts of the lens galaxy at a redshift $z_l \simeq 0.04$ with
axisymmetric dynamical models. The resulting intrinsic mass
distribution agrees well with that inferred from the lens model that
fits the four quasar image positions and relative fluxes. Furthermore,
we compare this total mass distribution with the stellar mass
distribution, based on an independent estimate of the stellar
mass-to-light ratio $\MLstar$, to place constraints on the dark matter
distribution in the inner parts of the lens galaxy.

In Section~\ref{sec:observations}, we briefly describe the Einstein
Cross, and we present the photometric and spectroscopic observations.
In Section~\ref{sec:analysis}, we extract the two-dimensional stellar
kinematics, describe the dynamical modeling method, and construct
lens models. From the latter we infer the total mass distribution,
which we compare in Section~\ref{sec:results} with the luminous mass
distribution inferred from the observed surface brightness. We then
build axisymmetric dynamical models, and compare the resulting
dynamical mass-to-light ratio with that inferred from the lens models,
and in turn with an estimate of the stellar mass-to-light ratio. In
Section~\ref{sec:discconcl} we discuss our findings and we summarize
our conclusions.

Throughout we adopt the \wmap\ cosmological parameters for the Hubble
constant, the matter density and the cosmological constant, of
respectively $H_0=73$\,\kmsM, $\Omega_M=0.24$ and $\Omega_L=0.76$
\citep[WMAP3;][]{2007ApJS..170..377S}.

\section{Observations and data reduction}
\label{sec:observations}

\subsection{The Einstein Cross}
\label{sec:einsteincross}

The Einstein Cross is the well-known gravitational lens system
QSO\,2237+0305 (RA: 22h\,40m\,30.3s, Dec:
$+03^\circ$\,21\arcmin\,31\arcsec, J2000). In this system, a distant
quasar at redshift $z_s = 1.695$ is lensed by the bulge of the
early-type spiral PGC\,069457 at $z_l = 0.0394$ (angular diameter
distance $D_l = 155$\,Mpc, $1\arcsec = 0.75$\,kpc), resulting in a
cross of four bright images separated by about 1.8\arcsec.

The Einstein Cross has long been the closest strong gravitational lens
system known, and has been very well studied since its discovery by
\cite{1985AJ.....90..691H}.  There is a wealth of ground- and
space-based imaging data at all wavelengths
\citep[e.g.][]{1996AJ....112..897F, 1998MNRAS.298.1223B,
  2000ApJ...545..657A, 2003ApJ...589..100D}. The resulting precise
measurements of the positions and relative fluxes of the quasar images
can be used to construct a detailed lens model.

In contrast, kinematic data of the lens galaxy is very scarce, with
only one measured central stellar velocity dispersion
\citep{1992ApJ...386L..43F} and two \HI\ rotation curve measurements
in the very outer parts \citep{1999MNRAS.309..641B}. There are several
previous integral-field studies of the Einstein Cross: \texttt{TIGER}:
\cite{1994A&A...282...11F}; \texttt{INTEGRAL}:
\cite{1998ApJ...503L..27M}; \texttt{CIRPASS}:
\cite{2004ApJ...607...43M}.  However, none of these studies were
concerned with the stellar kinematics of the lens galaxy, but instead
investigated the quasar spectra.

\subsection{Imaging}
\label{sec:imaging}

\begin{figure*}
  \includegraphics[width=0.330\textwidth]{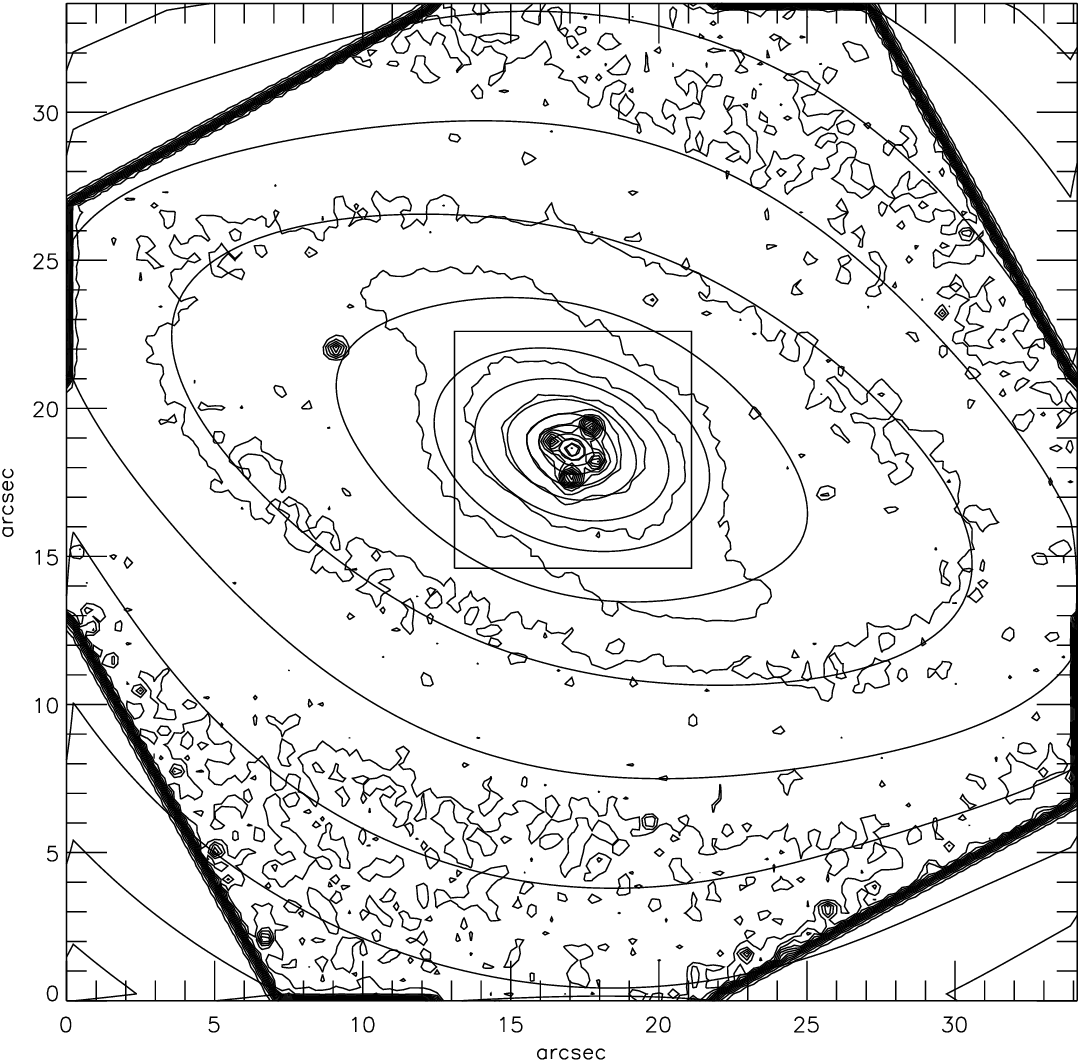}
  \hfill
  \includegraphics[width=0.325\textwidth]{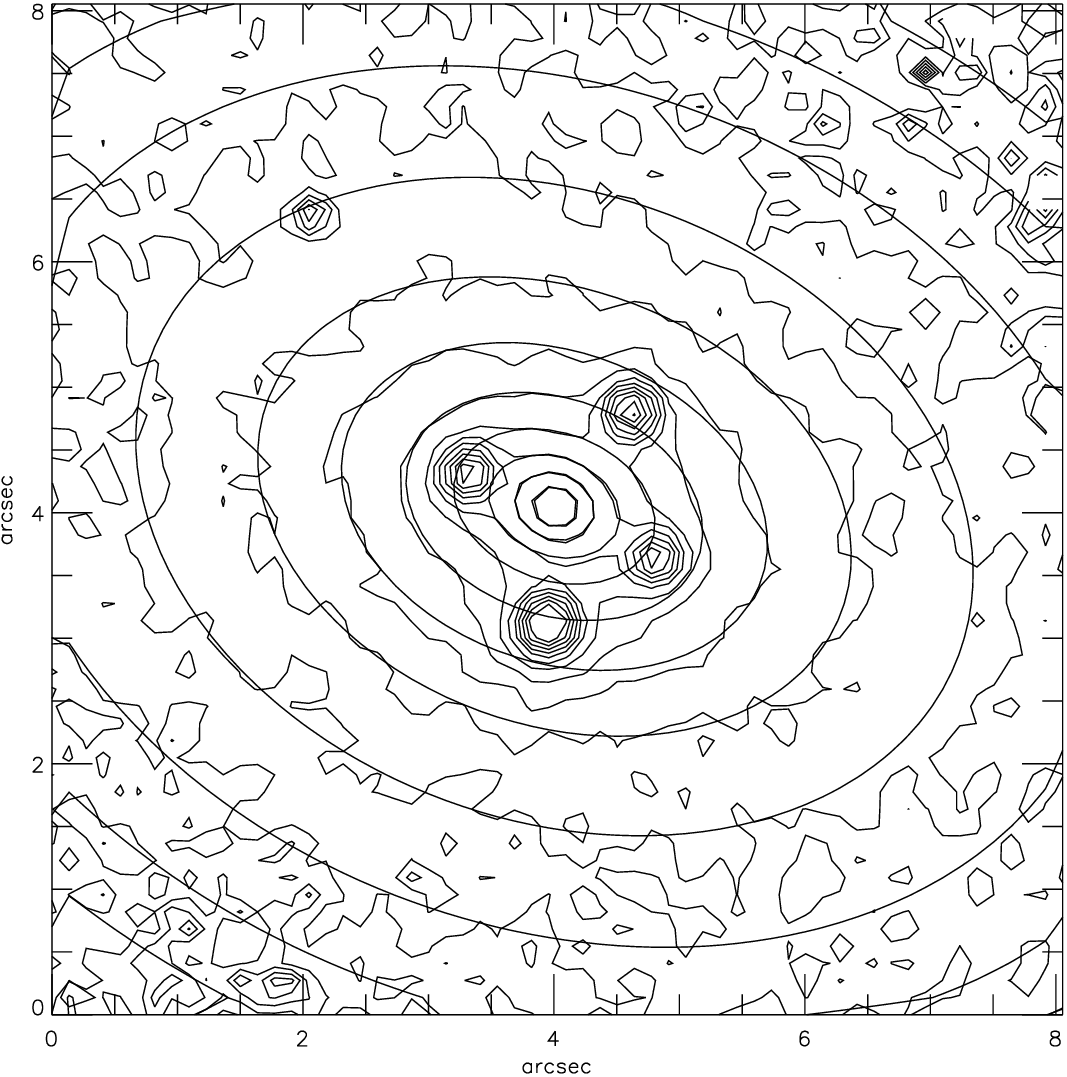}
  \hfill \includegraphics[width=0.330\textwidth]{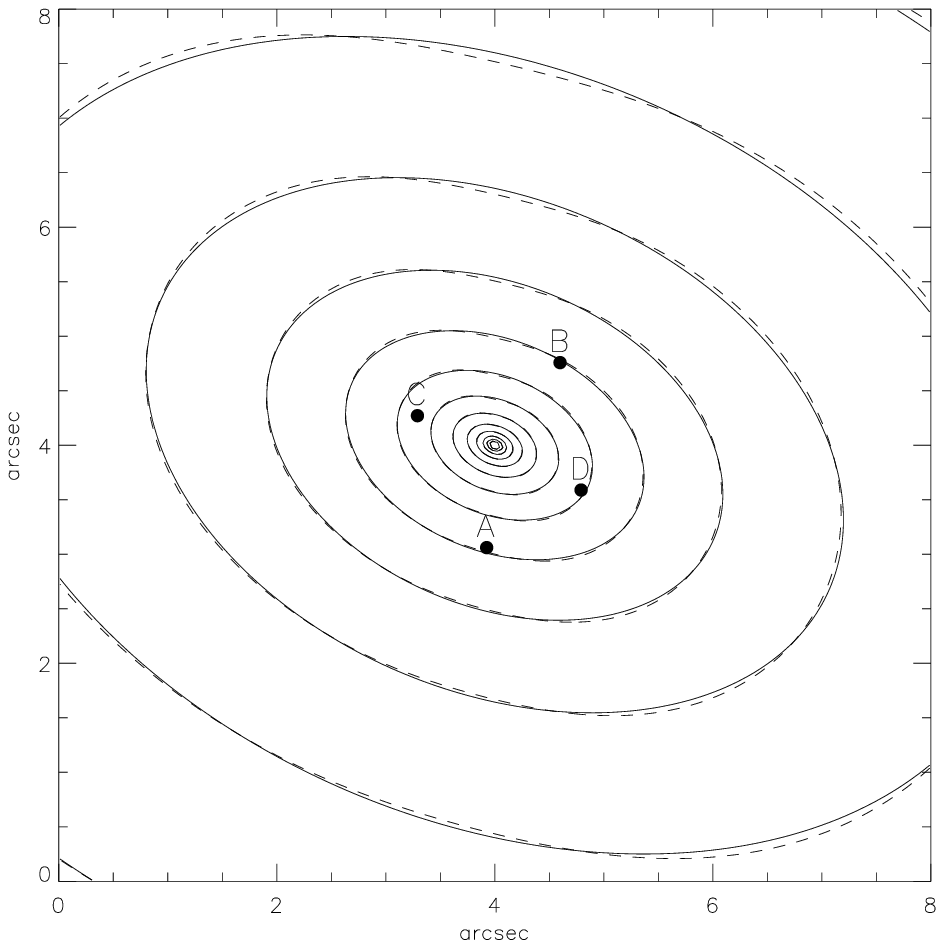}
  \caption{Surface brightness as observed with \hst\ (left and middle
    panel) and surface mass density of the overall best-fit lens model
    (right panel) of the lens galaxy in the Einstein Cross. \emph{Left
      panel}: the contours of the WFPC2/F555W $V$-band image reveal
    clearly the bulge, spiral arms and bar embedded in the large-scale
    disk of this early-type spiral galaxy.  The ellipticity measured
    from the MGE fit (contours) is used to estimate the inclination.
    \emph{Middle panel}: the central 8\arcsec$\times$8\arcsec\ of the
    WFPC2/F814W $I$-band image, of which the MGE fit is used to
    construct the (stellar) luminosity density model of the lens
    galaxy.  We use the $I$-band image instead of the longer exposed
    $V$-band image as it tracers better the old stellar population and
    is less sensitive to extinction and reddening. The four quasar
    images are masked out during the MGE fit. In both WFPC2 images the
    contours are in steps of $0.5$\,mag/arcsec$^2$. The images are
    rotated such that North is up and East is to the left.
    \emph{Right panel}: The scale-free lens model with slope
    $\alpha=1.0$ (dashed contours) fits the positions and relative
    fluxes of the quasar images, indicated by the filled circles.
    Superposed is the MGE fit (solid contours). }
  \label{fig:mgesurf}
\end{figure*}

We use two WFPC2 images retrieved from the \hst-archive (F555W
$V$-band image, 1600 seconds, PI: Westphal, and F814W $I$-band image,
120 seconds, PI: Kochanek; see left and middle panels of
Fig.~\ref{fig:mgesurf}) to determine the luminosity density of the
lens galaxy. We correct the $I$-band image for a Galactic extinction
of $E(B-V)=0.071$ \citep{1998ApJ...500..525S}, and we convert to solar
units using the WFPC2 calibration of \cite{2000PASP..112.1397D}, while
assuming an absolute $I$-band magnitude for the Sun of 4.08 mag
\citep[Table~2 of ][]{1998gaas.book.....B}.  From a de Vaucouleurs
$R^{1/4}$ profile fit to the $I$-band photometry in the inner
bulge-dominated region, we obtain an effective radius $\Reff \simeq
6$\arcsec, which is consistent with previous measurements
\citep[e.g.][]{1991AJ....102..454R}.

For the construction of the lens model, we use the accurate positions
of the quasar images from the website of the \castles\ 
survey\footnote{http://cfa-www.harvard.edu/castles/} based on
\textit{Hubble Space Telescope} (\hst) imaging. Although also optical
flux ratios are given on this website, we use the radio fluxes
provided by \cite{1996AJ....112..897F}, because they are in general
(much) less affected by differential extinction or microlensing.

\subsection{Integral-field spectroscopy}
\label{sec:integralfield}

Observations of the Einstein Cross lens system were carried out using
the integral-field unit of the \gmos-North spectrograph
\citep{2003SPIE.4841.1750M, 2004PASP..116..425H} on July
17$^\mathrm{th}$ and August 1$^\mathrm{st}$ 2005 as part of the
program GN-2005A-DD-7; the seeing was about $0.5$\arcsec. The data
were obtained using the IFU two-slit mode that provides a
field-of-view of 5\arcsec$\times$7\arcsec.  An array of 1500 hexagonal
lenslets, of which 500 are located 1\arcmin\ away from the main field
to be used for sky subtraction, sets the 0\farcs2 spatial sampling.
Eight individual science exposures of 1895 seconds each were obtained
during the two nights, resulting in a total on-source integration time
of $\simeq 4$ hours.  An offset of 0\farcs3 was introduced between
exposures to avoid bad CCD regions or lost fibers.  The R400-G5305
grating in combination with the CaT-G0309 filter was used to cover a
wavelength range of 7800--9200\,\AA\ with a FWHM spectral resolution
of 2.8\,\AA.

For the data reduction we used an updated version of the officially
distributed Gemini IRAF\footnote{IRAF is distributed by the National
  Optical Astronomy Observatories, which are operated by the
  Association of Universities for Research in Astronomy, Inc., under
  cooperative agreement with the National Science Foundation.}
package. We applied bias subtraction, flat-fielding and cosmic ray
rejection \citep[using the \emph{L.A.\ Cosmic} algorithm
by][]{2001PASP..113.1420V} to each science exposure, and wavelength
calibration after the extraction of the data.  A careful flat-fielding
procedure, crucial for proper removal of fringes in the spectral
direction, was carried out using the Quartz halogen (QH) lamp.
Wavelength calibration was performed with CuAr lamp exposures taken
before each science exposure. At the observed wavelengths, a complex
spectrum of H$_2$O absorption features overlaps with the CaT lines we
are interested in.  We constructed a correction spectrum from
observations of the white dwarf star Wolf1346, taken with the same
instrumental setup as our science frames.  We checked the range of
fiber-to-fiber variations of the spectral resolution of the instrument
by measuring the width of the sky lines in each fiber, resulting in
the nominal FWHM value of $2.8 \pm 0.2$\,\AA\ for most of the data
cubes.  In order to combine all the data, we homogenized the science
frames by convolving all spectra to an instrumental FWHM resolution of
3.1\,\AA\ (or $\sigma_\mathrm{instr} \simeq 44.2$\,\kms) and
resampling the spectra to the same range and sampling in wavelength.
After interpolating each science frame to a common spatial grid,
taking into account the small spatial offsets applied during the
observations, we sum the spectra sharing the same position in the sky
to produce the final merged data cube.

The significant contribution from the sky lines to the overall
spectrum of the lens galaxy made the sky subtraction the most
challenging step in the data reduction process. Although the scatter
in the instrumental resolution was small within the individual science
frames, the combined effect of other steps in the data reduction (e.g.
fringing, telluric absorption, resampling of data in wavelength),
introduced alterations in the shape of the sky lines (from the sky
fibers) that were not equally reproduced in the science fibers.  As a
consequence any attempt to use the sky from the sky fibers resulted in
serious systematic effects (i.e.\ P-Cygni like residuals) in the galaxy
spectra underneath. In order to minimize these effects, we chose to
extract the sky spectra from the outer most regions of our fully
merged science data cube, where no galaxy light was appreciable. The
sky was then subtracted optimally as described below.

\section{Analysis}
\label{sec:analysis}

\subsection{Two-dimensional stellar kinematics}
\label{sec:stellarkin}

\begin{figure}
  \begin{center}
    \includegraphics[width=1.0\columnwidth]{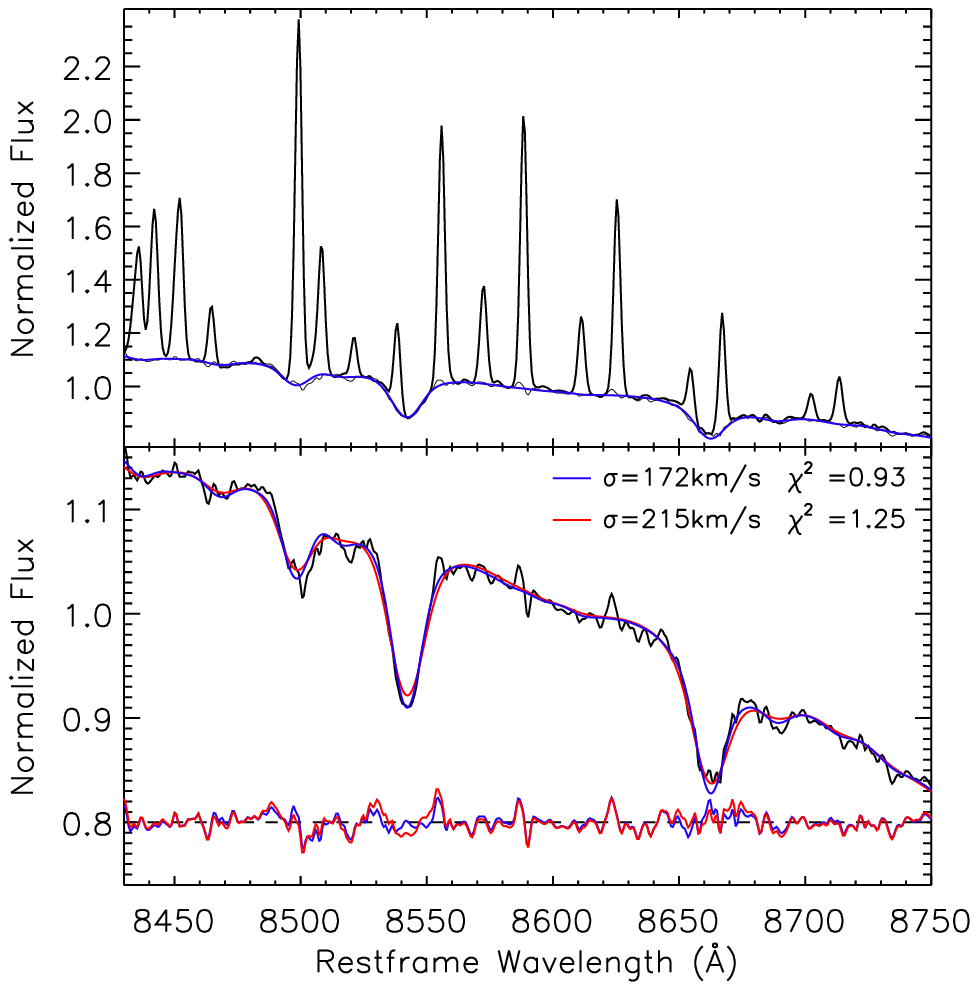}
  \end{center}
  \caption{Spectrum from the center of the lens galaxy. \emph{Upper
      panel:} The observed spectrum (thick black curve) clearly shows
    the contamination by the sky emissions lines. The sky-subtracted
    galaxy spectrum (thin black curve) shows the Ca~II triplet
    absorption lines which are nicely fitted by a composite of stellar
    population models (blue curve). \emph{Lower panel:} The same
    sky-subtracted galaxy spectrum (black curve) and composite stellar
    population model (blue curve) are shown. In addition to the this
    model with a best-fit velocity dispersion of 172\,\kms, another
    model (red curve) is shown with the velocity dispersion fixed to
    the 215\,\kms\ as measured by \cite{1992ApJ...386L..43F}.  The
    residuals at the bottom and the quoted reduced $\chi^2$-values
    show that 172\,\kms provides a better fit than 215\,\kms,
    particularly in fitting the strongest of the three absorption
    lines. Note that this spectrum is extracted from the same central
    aperture of 0.7\arcsec$\times$0.4\arcsec\ (at a position angle of
    39\dgr) as used by \cite{1992ApJ...386L..43F}, and that the seeing
    conditions were similar. Throughout we use instead a circular
    aperture with radius 1\arcsec\ to derive the central velocity
    dispersion $\sigma_0 = 170$\,\kms.}
  \label{fig:censpec}
\end{figure}

\begin{figure}
  \begin{center}
    \includegraphics[width=1.0\columnwidth]{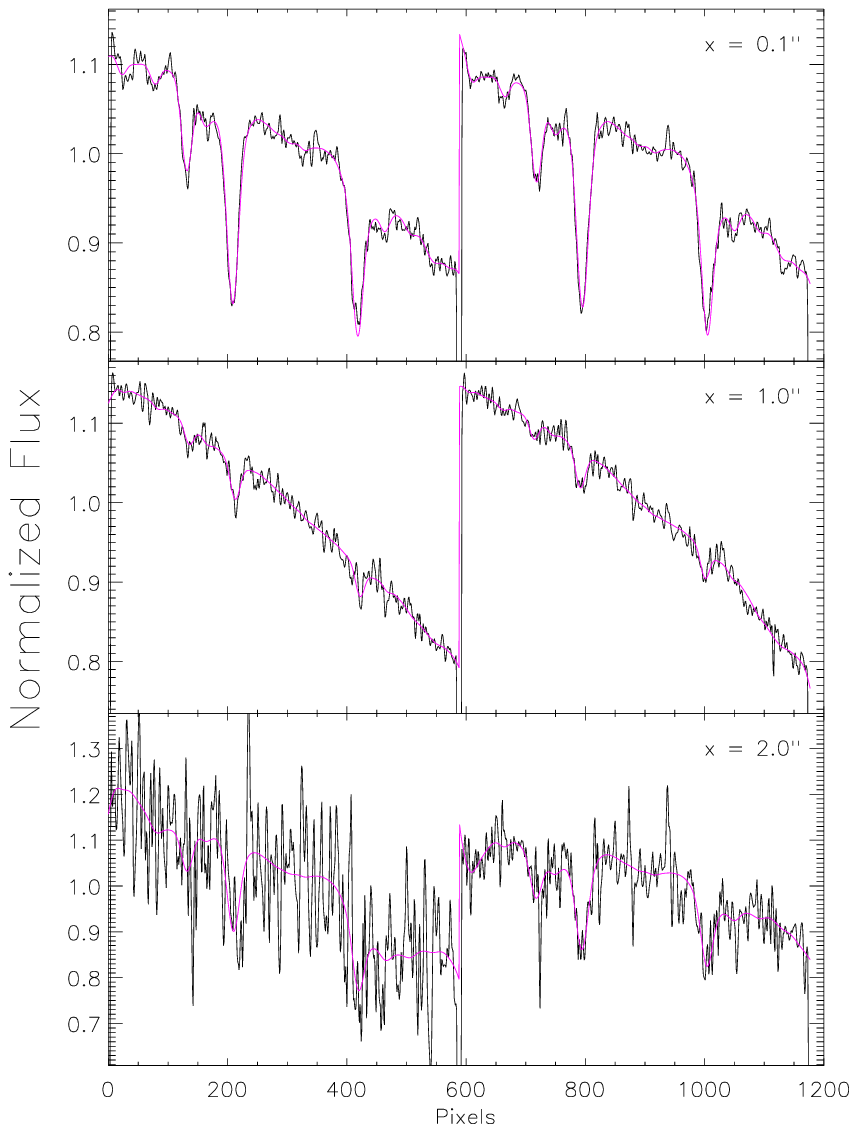}
  \end{center}
  \caption{Symmetrized fits to (sky-subtracted) spectra at three
    different positions along the major axis. Each panel shows the two
    spectra on opposite positions from the center at a distance of
    0.1\arcsec\ (top), 1.0\arcsec\ (middle), and 2.0\arcsec\
    (bottom). Assuming (axi)symmetry with respect to the minor axis,
    the same composite stellar population model is fitted to both
    sides, apart from an opposite sign in the centroid velocity. This
    symmetrized fitting not only reduces impact of systematic effects
    in the data, but also increases the effective
    signal-to-noise. This can significantly improve the extracted
    kinematics, as can already be seen from the bottom panel where the
    left spectrum is of lesser quality than the right spectrum, and
    from the full kinematic maps in Fig.~\ref{fig:eckinmaps}. In the
    middle panel, the contribution (in this case contamination) from
    the quasar source is clearly visible, and even though the fits to
    the Ca~II triplet still look reasonable the corresponding position
    is excluded.}
  \label{fig:eckinspec}
\end{figure}

\begin{figure*}
  \begin{center}
    \includegraphics[width=1.0\textwidth]{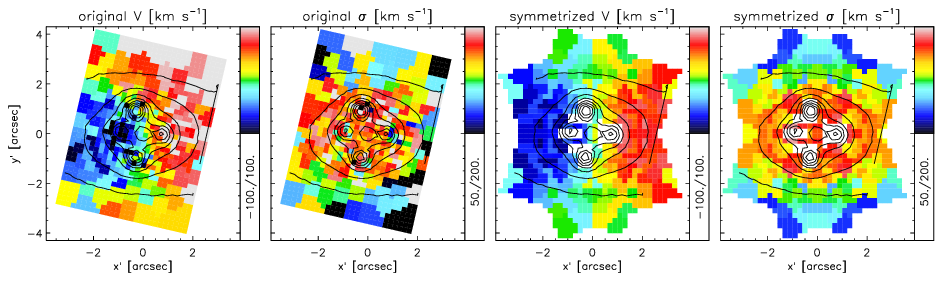}
  \end{center}
  \caption{Mean line-of-sight velocity $V$ and velocity dispersion
    $\sigma$ maps of the lens galaxy in the Einstein Cross as measured
    from observations with the integral-field spectrograph \gmos\ on
    Gemini-North. At the right side of each map, the (linear) scale in
    \kms\ is indicated by the color bar, and the limits are given
    below. The third and fourth panel show the $V$ and $\sigma$ maps
    after symmetrization assuming oblate axisymmetry (see
    \S~\ref{sec:stellarkin}). The overplotted (black) contours show
    the isophotes of the image reconstruction from collapsing the
    unsymmetrized data cube, highlighting the regions excluded in the
    symmetrized kinematics due to contamination from the quasar
    images. The $V$ map shows clear and regular rotation around the
    (vertically aligned) short axis of the bulge. The $\sigma$ map
    shows a weak gradient decreasing towards the edge of the field.}
  \label{fig:eckinmaps}
\end{figure*}

To measure reliable stellar kinematics we first co-add spectra using
the adaptive spatial two dimensional binning scheme of
\cite{2003MNRAS.342..345C} to obtain in each resulting (Voronoi) bin a
minimum signal-to-noise ratio (S/N). The resulting spectra are
spectrally rebinned to steps of constant $\ln\lambda$, equivalent to a
sampling of 24\,\kms\ per pixel.  We adopt the single stellar
population (SSP) models of \cite{2003MNRAS.340.1317V} as spectral
templates, which are convolved to have the same effective instrumental
broadening as the \gmos\ data.  Next, using an updated version (v4.2)
of the penalized pixel-fitting (pPXF) algorithm of
\cite{2004PASP..116..138C}, a non-negative linear combination of these
templates is convolved with a Gaussian line-of-sight velocity
distribution (LOSVD) and fit to each observed spectrum, to derive the
mean line-of-sight velocity $V$ and velocity dispersion $\sigma$ per
bin on the plane of the sky. This version of the code allows for the
inclusion of a sky spectrum into the set of stellar templates, and
determines the optimal scaling of the sky to the overall spectrum
\citep[see also][]{2009MNRAS.398..561W}. We used this procedure to
\emph{clean} the individual spectra in our merged data cube, before we
derived the final stellar kinematics.

As an example of how this routine performed, we show in the top panel
of Fig.~\ref{fig:censpec} the observed spectrum from the center of the
lens galaxy (thick black curve), and the cleaned galaxy spectrum (thin
black curve) fitted by a composite of stellar population models (blue
curve). The sky emission-lines can severely affect the Ca~II triplet
absorption lines that we use to extract the stellar kinematics,
especially at larger radii where the the relative contribution from
the sky is more significant. In particular, in our experiments, a poor
sky subtraction caused significant systematic effects in the map of
$\sigma$ at the receding side of the galaxy, where one of the
absorption lines is redshifted onto (the residual of) a sky emission
line.  The map of $V$, however, was (nearly) unaffected, showing
regular rotation with a straight zero-velocity curve that coincides
with the minor axis of the (bulge) photometry. These tests already
show that, at least in the inner parts of the lens galaxy, both the
photometry and kinematics are consistent with axisymmetry.

If we now \emph{assume} this symmetry from the start and apply it
during the extracting of the stellar kinematics, we can not only
increase the S/N, but at the same time also suppress or even remove
systematic effects. To do this, we make use of a built-in
functionality in the pPXF algorithm that allows diametrically opposed
spectra to be fitted simultaneously. The symmetry imposed by this
technique can be either mirror-symmetric or point-symmetric,
corresponding to the behavior of the LOSVD in axisymmetric or triaxial
stellar systems, respectively. In practice, this is done following the
method outlined by \cite{1992MNRAS.254..389R}, whereby the two spectra
from opposed positions in the galaxy are laid end-to-end to create a
single vector. The template library is also doubled in length, where
the second half is convolved by the symmetric counterpart of the
broadening function used for the first half (simply reflecting the
broadening kernel around the systemic velocity). The two galaxy
spectra are therefore fitted simultaneously, with the implied
assumption of symmetry. 

Performing a symmetric extraction of the kinematics presents a number
of advantages, and has been used by numerous authors not just as a
quality-check \citep[e.g.][]{1992MNRAS.254..389R}, but also when the
assumption of symmetry is an integral part of the analysis
\citep[e.g.][]{2003ApJ...583...92G}. The two main advantages for the
case presented here are in reducing the impact of systematic effects
in the data (including sky subtraction) and increasing the effective
spatial resolution. The latter is possible since under the assumption
of axisymmetry, all moments of the LOSVD are symmetric about the
galaxy major axis. We can therefore `fold' the data cube about the
major axis, combining spectra from the two hemispheres, which
increases the effective S/N within a given spatial element, and
therefore reduces the required spatial bin size.

We interpolate the merged \gmos\ data cube onto a new spatial grid
with $x'$ and $y'$ coordinates parallel to the major and minor
photometric axes respectively. We then fold the data-cube about the
major axis by summing together spatial pixels (`spaxels') with equal
$(x,|y|)$ positions. Before spatially binning the spectra to a minimum
S/N, the folded data cube is folded again, into a single quadrant by
combining spaxels with equal $(|x|,y)$ positions. The Voronoi bins are
derived in this quadrant, and the resulting bin centers are mirrored
around $x'=0$. The spectra are combined according to these Voronoi
bins to ensure symmetry around the minor axis. At the field perimeter,
it is generally not the case that both sides of the galaxy are sampled
at the exact same positions, due to the alignment of the \gmos\ field
with respect to the galaxy's principle axes. This results in some bins
where the S/N of the opposed spectra is far from equal, or in some
cases where observations are only present on one side of the galaxy.
The former is taken into account by consideration of the error
spectrum of each bin. For the latter, the bin is fitted as a single
spectrum, giving kinematic values only at that position, without a
mirrored counterpart. 
Fig.~\ref{fig:eckinspec} shows how the (axi)symmetric kinematic
extraction performed at three different position along the major axis.

From Monte-Carlo simulations for data of this spectral range and
resolution, a minimum S/N of 20 is deemed optimal. This results in an
average error in both $V$ and $\sigma$ of about $9$\,\kms, while still
preserving the spatial resolution of the data.  However, towards the
edge of the \gmos\ field, the errors increase to $\sim 15$\,\kms\ in
$V$ and $\sim 20$\,\kms\ in $\sigma$. The final symmetrically binned
data cube consists of $4 \times 71 = 284$ bins, with the largest
containing $19$ individual 0.2\arcsec\ spaxels.  The resulting
(original and symmetrized) maps of $V$ and $\sigma$ are presented in
Fig.~\ref{fig:eckinmaps}, with superposed contours of the integrated flux
from the data cube. The velocity map shows regular rotation with
amplitude of $V$ up to $\sim 100$\,\kms. The dispersion map shows
somewhat higher values of $\sigma$ at the position of the quasar
images. This is caused by imperfect subtraction of the quasar
continuum, which acts to strongly dilute the absorption lines and make
them appear broader.

In our stellar template fit to the observed spectra, we include
additive polynomials up to 10th order to account for remaining
systematic effects in the data, e.g. due to imperfect flat
fielding. At the same time, we use these additive components to mimic
the contaminating contribution of the quasar images, which impart a
strong featureless continuum on top of the galaxy's integrated stellar
spectrum. This method nicely removes nearly all contribution from the
quasar, except where the quasar images are the brightest, in which
case $\sigma$ cannot be reliable measured. 
The middle panel of Fig.~\ref{fig:eckinspec} provides an example of
such a case in which the spectra come from the opposite positions
1.0\arcsec\ from the center along the major axis, right in the middle
of two quasar images. Even though the composite stellar population
model fits to the Ca~II triplet still look reasonable, we exclude
these regions when we fit dynamical models to the $V$ and $\sigma$
maps.
However, between the quasar images, the spectra are dominated by the
stellar light of the galaxy, giving a reliable central ($R<1$\arcsec)
dispersion $\sigma_0 \simeq 170 \pm 9$\,\kms.
Likewise, at the edges of the field, the quasar contamination is
small, providing a reliable gradient in the velocity dispersion.

Comparing in Fig.~\ref{fig:eckinmaps} the symmetrized stellar
kinematics with what is obtained without imposing axisymmetry, there
is a marked reduction of noise and systematic effects, especially in
the velocity dispersion map. However, the main features of regular
rotation, decreasing $\sigma$ at larger radii, as well as
corresponding velocity and dispersion amplitudes, are present in both
approaches, albeit with larger uncertainty and systematics in the
non-symmetrized case. Given the various advantages, we consider
hereafter the symmetrically extracted kinematics.

\subsection{Dynamical models}
\label{sec:dynamicalmodel}

As mentioned above, both the photometry and (unsymmetrized) kinematics
in the inner parts of the lens galaxy are consistent with axisymmetry.
Hence, in constructing dynamical models we consider an axisymmetric
stellar system in which both the potential $\Phi(R,z)$ and
distribution function (DF) are independent of azimuth $\phi$ and time.
By Jeans' (\citeyear{1915MNRAS..76...70J}) theorem the DF only depends
on the isolating integrals of motion: $f(E,L_z,I_3)$, with energy $E =
(v_R^2+v_\phi^2+v_z^2)/2 + \Phi(R,z)$, angular momentum $L_z =
Rv_\phi$ parallel to the symmetry $z$-axis, and a third integral $I_3$
for which in general no explicit expression is known. However,
usually\footnote{If resonances are present, $I_3$ may loose this
  symmetry.} $I_3$ is invariant under the change $(v_R,v_z) \to
(-v_R,-v_z)$. This implies that the mean velocity is in the azimuthal
direction ($\overline{v_R}=\overline{v_z}=0$) and the velocity
ellipsoid is aligned with the rotation direction ($\overline{v_R
  v_\phi} = \overline{v_\phi v_z} = 0$).

\cite{1979ApJ...232..236S} introduced a method that sidesteps our
ignorance about the non-classical integrals of motion. It finds the
set of weights of orbits computed in an arbitrary gravitational
potential that best reproduces all available photometric and kinematic
data at the same time. The method has proved to be powerful in
building detailed spherical and axisymmetric models of nearby galaxies
\citep[e.g.][]{1997ApJ...488..702R, 1998ApJ...493..613V,
  2003ApJ...583...92G, 2004ApJ...602...66V, 2006MNRAS.366.1126C,
  2007MNRAS.382..657T} as well as globular clusters
\citep{2006A&A...445..513V, 2006ApJ...641..852V}, and since recently
also triaxial models \citep{2008MNRAS.385..614V, 2008MNRAS.385..647V}.
However, in all cases the stellar systems are significant closer than
the lens galaxy in the Einstein Cross, allowing for (even) more and
higher-order stellar kinematic measurements necessary to constrain the
large freedom in this general modeling method. Instead we construct
simpler, but still realistic dynamical models based on the solution of
the axisymmetric Jeans equations.

When we multiply the collisionless Boltzmann equation in cylindrical
coordinates by respectively $v_R$ and $v_z$ and integrate over all
velocities, we obtain the two Jeans equations \citep[see
also][eq.~4-29]{BT87}
\begin{eqnarray}
  \label{eq:cylcbevR}
  \frac{\partial(R\nu\overline{v_R^2})}{\partial R}
  + R\frac{\partial(\nu\overline{v_R v_z})}{\partial z}
  - \nu\overline{v_\phi^2} 
  + R\nu\frac{\partial \Phi}{\partial R} 
  & = & 0,
  \\
  \label{eq:cylcbevz}
  \frac{\partial(R\nu\overline{v_R v_z})}{\partial R}
  + R\frac{\partial(\nu\overline{v_z^2})}{\partial z}
  + R \nu \frac{\partial \Phi}{\partial z} 
  & = & 0,
\end{eqnarray}
where $\nu(R,z)$ is the intrinsic luminosity density. Due to the
assumed axisymmetry, all terms in the third Jeans equation, that
follows from multiplying by $v_\phi$, vanish.

We are thus left with four unknown second order velocity moments
$\overline{v_R^2}$, $\overline{v_z^2}$, $\overline{v_\phi^2}$ and
$\overline{v_R v_z}$ and only two equations. This means we have to
make assumptions about the velocity anisotropy, or in other words the
shape and alignment of the velocity ellipsoid. In case the velocity
ellipsoid is aligned with the cylindrical $(R,\phi,z)$ coordinate
system $\overline{v_R v_z} = 0$, so that we can readily solve
equation~\eqref{eq:cylcbevz} for $\overline{v_z^2}$. If we next assume
a constant flattening of the velocity ellipsoid in the meridional
plane, we can write $\overline{v_R^2} = \overline{v_z^2}/(1-\beta_z)$
and solve equation~\eqref{eq:cylcbevR} for $\overline{v_\phi^2}$. This
assumption provides in general a good description for the kinematics
of real disk galaxies \citep{2008MNRAS.390...71C}, such as the lens
galaxy in the Einstein Cross, which is an early-type spiral galaxy.
When $\beta_z = 0$, the velocity distribution is isotropic in the
meridional plane, corresponding to the well-known case of a
two-integral DF $f(E,L_z)$ \citep[e.g.][]{1962MNRAS.123..447L,
  1977AJ.....82..271H}.

Knowing the intrinsic second-order velocity moments, the line-of-sight
second-order velocity moment for a stellar system viewed at an
inclination $i > 0$ away from the $z$-axis follows as
\begin{eqnarray}
  \label{eq:cylvlosmge}
  \overline{v_\mathrm{los}^2} & = & \frac{1}{I(x',y')}
  \int_{-\infty}^{+\infty} \nu \biggl[
    \left(\overline{v_R^2}\sin^2\phi +
      \overline{v_\phi^2}\cos^2\phi\right) \sin^2i
  \biggr.
  \nonumber \\ 
  && 
  \biggl. 
  + \overline{v_z^2} \cos^2i 
  - \overline{v_R v_z}\sin\phi\sin(2i) 
  \biggr] \, \du z',
\end{eqnarray}
where $I(x',y')$ is the (observed) surface brightness with the
$x'$-axis along the projected major axis. For each position $(x',y')$
on the sky-plane, $\overline{v_\mathrm{los}^2}$ yields a prediction of
the second velocity moment $V^2 + \sigma^2$, which is a combination of
the (observed) mean line-of-sight velocity $V$ and velocity dispersion
$\sigma$.

Under the above assumptions, besides the anisotropy parameter $\beta_z$
(and possibly the inclination $i$), the only unknown quantity is the
gravitational potential.
In other words, once the gravitational potential is known, the second
velocity moment, together with the assumed velocity anisotropy, define
which stellar orbits are present, except for their sense of rotation.
Almost any velocity field $V$ can then be reproduced by arranging the
sense of rotation of the individual orbits, without any change to the
gravitational potential. The only limit on $V$ is when all the orbits
rotate in the same direction, but this limit is not very useful since
it is generally much larger than the observed velocity in galaxies.
This implies that, unless further assumptions are made, the observed
velocity $V$ is virtually useless for determining the gravitational
potential.
Henceforth, we use the combined $V^2 + \sigma^2$ to constrain the
gravitational potential.

The gravitational potential is via Poisson's equation related to the
total mass density $\rho(R,z)$. 
We may estimate the latter from the intrinsic luminosity density
$\nu(R,z)$, derived from deprojecting the observed surface brightness
$I(x',y')$, once we know the total mass-to-light ratio $\MLtot$. It is
common in dynamical studies of the inner parts of galaxies to consider
$\MLtot$ as an additional parameter and to assume its value to be
constant, i.e., mass follows light
\citep[e.g.][]{2006MNRAS.366.1126C}. Since $\MLtot$ may be larger than
the stellar mass-to-light ratio $\MLstar$, this still allows for
possible dark matter contribution, but with constant fraction.
As an alternative to these constant-$\MLtot$ models, we
also construct dynamical models in which we use the strong
gravitational lensing to constrain the gravitational potential. In
\S~\ref{sec:lensmodel} below, we use the positions and relative fluxes
of the quasar images to estimate the surface mass density
$\Sigma(x',y')$, which we deproject to obtain $\rho(R,z)$, so that
$\MLtot$ is not anymore a free parameter. 
For both type of dynamical models, the best-fits to the stellar
kinematic data are shown in Fig.~\ref{fig:jeansmodel}, and discussed
further in \S~\ref{sec:axijeansmodels} below.

We use the Multi-Gaussian Expansion method
\citep[MGE;][]{1992A&A...253..366M,1994A&A...285..723E} to
parameterize both the observed surface brightness $I(x',y')$ as well
as the estimated surface mass density $\Sigma(x',y')$ as a set of
two-dimensional Gaussians. Even though Gaussians do not form a
complete set of functions, in general surface density distributions
are accurately reproduced, including deviations from an elliptical
distribution and ellipticity variations with radius. Representing also
the point-spread function (PSF) by a sum of Gaussians, the convolution
with the PSF becomes straightforward.  Moreover, the
MGE-parameterization has the advantage that the deprojection can be
performed analytically once the viewing angle(s) are given. Also many
intrinsic quantities such as the potential can be calculated by means
of simple one-dimensional integrals. 
Similarly, the calculation of $\overline{v_\mathrm{los}^2}$ in
equation~\eqref{eq:cylvlosmge} reduces from the (numerical) evaluation
of in general a triple integral to a straightforward single integral
\citep[for further details and a comparison with more general
dynamical models see][]{2008MNRAS.390...71C}. 
The latter integral is given in Appendix~\ref{sec:appmge}, together
with expressions for $\nu(R,z)$ and $\Phi(R,z)$ for the
MGE-parameterization of $I(x',y')$ and $\Sigma(x',y')$ as given in
Table~\ref{tab:mgepar}, and further discussed in
\S~\ref{sec:massdistr} below.
We used the publically available Jeans Anisotropic MGE (JAM)
implementation\footnote{Available from http://purl.org/cappellari/idl}
to compute $\overline{v_\mathrm{los}^2}$, also including PSF
convolution and sampling over the GMOS lenslets.

\subsection{Lens models}
\label{sec:lensmodel}

\begin{figure*}
  \begin{center}
    \includegraphics[width=1.0\textwidth]{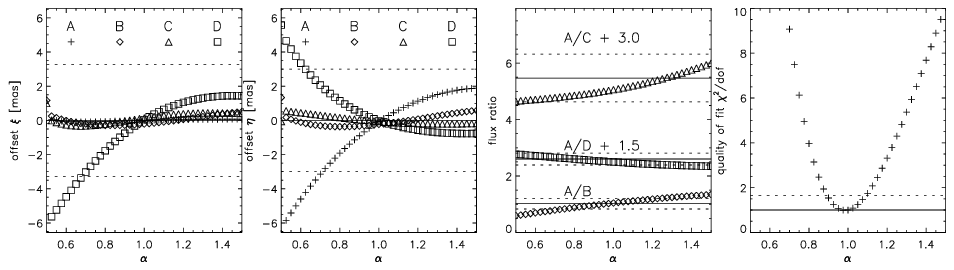}
  \end{center}
  \caption{Lens model with $n=4$ Fourier terms (see
    \S~\ref{sec:lensmodel} for details) fitted to the position of the
    quasar source (first two panels, in milliarcseconds) and to the
    observed radio flux ratios of the quasar images (third panel).
    Over the full range of slopes $0.5<\alpha<1.5$, the predicted
    source positions from each quasar image A through D (as indicated
    by the different symbols) fall within the observed error (dotted
    lines).  However, the recovery of the observed flux ratios (solid
    lines) within the observed errors (dotted lines) places
    constraints on the slope.  The minimum reduced $\chi^2$ value of
    about unity in the fourth panel shows that for $\alpha \simeq 1.0$
    a good overall best-fit model is obtained, with corresponding
    $99$\,\%-confidence interval (dotted line) of $0.9 \lesssim \alpha
    \lesssim 1.1$.}
  \label{fig:ecfit}
\end{figure*}

\begin{figure*}
  \begin{center}
    \includegraphics[width=1.0\textwidth]{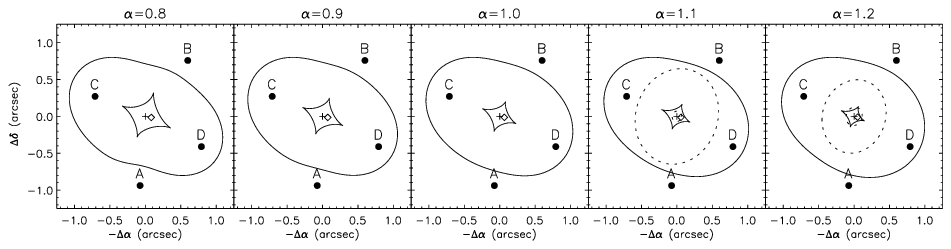}
  \end{center}
  \caption{ Best-fit lens models for five different values of the
    slope $\alpha$ indicated at the top of each panel (see
    Fig.~\ref{fig:ecfit} for corresponding quality of the fit). The
    filled circles indicate the positions of the four quasar images,
    the cross represents the center of the lens galaxy, and the
    diamond shows the best-fit position of the quasar source. The
    smooth and pinched solid curves show respectively the tangential
    critical curve and radial caustic.  The dashed curves are the
    radial counterparts when they exist (for $\alpha > 1.0$).}
  \label{fig:ecmod}
\end{figure*}

\begin{table*}
\begin{center}
\caption{Lens model parameters\label{tab:ecmod}}
\begin{tabular}{c*{13}{c}}
\hline
\hline
$\alpha$ & $\xi$ & $\eta$ 
& $c_0$ & $c_2$ & $c_3$ & $c_4$ & $s_2$ & $s_3$ & $s_4$ & $\Rein$ & $\Mein$ & $\Mc$\\
\hline
0.8 &  0.0829 & -0.0148 &  2.1615 & -0.0588 &  0.0000 &  0.0018 &  0.0589 &  0.0021 &  0.0029 & 0.8858 & 1.5413 & 1.5998 \\
0.9 &  0.0738 & -0.0141 &  1.9457 & -0.0493 &  0.0001 &  0.0011 &  0.0499 &  0.0005 &  0.0019 & 0.8862 & 1.5426 & 1.5845 \\
1.0 &  0.0651 & -0.0132 &  1.7733 & -0.0415 &  0.0002 &  0.0006 &  0.0423 & -0.0007 &  0.0012 & 0.8866 & 1.5441 & 1.5756 \\
1.1 &  0.0569 & -0.0121 &  1.6325 & -0.0350 &  0.0004 &  0.0003 &  0.0357 & -0.0014 &  0.0008 & 0.8872 & 1.5460 & 1.5708 \\
1.2 &  0.0490 & -0.0110 &  1.5153 & -0.0293 &  0.0005 &  0.0001 &  0.0300 & -0.0020 &  0.0006 & 0.8878 & 1.5481 & 1.5688 \\
\hline
\end{tabular}
\tablecomments{Parameters of the lens models shown in
  Fig.~\ref{fig:ecmod}. For each slope $\alpha$, the best-fit quasar
  source position $(\xi,\eta)$ and Fourier coefficients $c_m$ and
  $s_m$ ($c_1=s_0=s_1=0$) are given, all in arcseconds.  The last
  three columns provide the Einstein radius (in arcsec), the projected
  mass within this radius and within the critical curve (both in
  $10^{10}$\,\Msun).}
\end{center}
\end{table*}

Under the thin-lens approximation, the gravitational lensing
properties of a galaxy are characterized by its potential projected
along the line-of-sight, also known as the deflection potential
$\phi(x',y')$. In case of a MGE parameterization of the surface mass
density of the galaxy, the projection along the line-of-sight of the
corresponding gravitational potential, or, alternatively, solving the
two-dimensional Poisson equation becomes straightforward. The
calculation of $\phi(x',y')$ and its derivatives reduce to the
(numerical) evaluation of a single integral, as shown in
Appendix~\ref{sec:appmgeaxilens} for an axisymmetric system. This MGE
approach thus provides a powerful technique to build general lens
models. However, since in this case the quasar images only provide a
few constraints within a limited radial range, we adopt instead a
simpler and less general approach.

We use the algorithm of \cite{2003MNRAS.345.1351E} to
construct a lens model that accurately fits the (optical) positions
and relative (radio) fluxes of the four quasar images in the Einstein
Cross. The deflection potential is assumed to be a scale-free function
$\phi(R,\theta) = R^\alpha F(\theta)$ of the polar
coordinates $R$ and $\theta$ in the lens sky-plane, with $0<\alpha<2$
for realistic models. The angular part $F(\theta)$ is expanded as a
Fourier series
\begin{equation}
\label{eq:eqFtheta}
    F(\theta) = \frac12\,c_0 + 
    \sum_{m=1}^n [c_m\cos(m\theta) +  s_m\sin(m\theta)].
\end{equation}
The positions $(x_i,y_i) = (R_i\cos\theta_i,R_i\sin\theta_i)$ of the
images are related to the position $(\xi,\eta)$ of the source by the
lens equation \citep[e.g.][]{1992grle.book.....S}
\begin{eqnarray}
  \xi  &=& x_i \left[ 1 - \left( \alpha F_i - \tan\theta_i \, F'_i
    \right) R_i^{\alpha-2} \right], 
  \nonumber \\ \label{eq:lenseq}
  \eta  &=& y_i \left[ 1 - \left( \alpha F_i + \cot\theta_i \, F'_i
  \right) R_i^{\alpha-2} \right], 
\end{eqnarray}
where $F_i \equiv F(\theta_i)$ and $F'_i$ denotes the derivative to
$\theta_i$. The flux ratios of the images follow from their
magnifications $\mu_i$, which are given by
\begin{multline}
  \label{eq:fluxeq}
  \left( p_i \mu_i \right)^{-1} =
  1 - \left( \alpha^2 F_i + F''_i \right) R_i^{\alpha-2} \\
  + (\alpha-1) \left[ \alpha^2 F_i^2 -(\alpha-1) {F'_i}^2 + \alpha F_i
    F''_i \right] R_i^{2(\alpha-2)},
\end{multline}
where $p_i$ is the parity of image $i$. 

Since the Fourier coefficients $c_1$ and $s_1$ correspond to a
displacement of the source position $(\xi,\eta)$, we set $c_1=s_1=0$
to remove this degeneracy. For a given slope $0<\alpha<2$, we are left
with $2n+1$ free parameters $(\xi,\eta,c_2,s_2,\dots,c_n,s_n)$ and
eleven constraints (four image positions and three flux ratios).

For each image position $(x_i,y_i)$, we obtain through the lens
equation~\eqref{eq:lenseq} a prediction of the source position
$(\tilde{\xi}_i,\tilde{\eta}_i)$, which we compare with the assumed
free parameter values $(\xi,\eta)$ for the source position by
evaluating
\begin{equation}
  \label{eq:Chi2src}
  \chi_\mathrm{src}^2 = 
  \sum_{i=1}^4 \left[
    \frac{(\xi-\tilde{\xi}_i)^2}{\Delta_{xx,i}^2} +
    2 \frac{(\xi-\tilde{\xi}_i)(\eta-\tilde{\eta}_i)}{\Delta_{xy,i}^2} +
    \frac{(\eta-\tilde{\eta}_i)^2}{\Delta_{yy,i}^2}
  \right],
\end{equation}
where the weights in the source position are given by 
\begin{equation}
  \label{eq:weightssrc}
  \frac{1}{\Delta_{st,i}^2} =
  \frac{m_{xs,i}m_{xt,i}}{\Delta_{x,i}^2} 
  + \frac{m_{ys,i}m_{yt,i}}{\Delta_{y,i}^2}.
\end{equation}
The conversion from the uncertainties in the observed image positions
$(\Delta_{x,i}, \Delta_{y,i})$ in the image plane to uncertainties in
the source position in the source plane involves the magnification
tensor\footnote{We implicitly assume here that the deviation between
  predicted and assumed source position is small enough that the
  magnification is nearly constant \citep[see
  also][]{2001astro.ph..2340K}.} with components
\begin{eqnarray}
  \label{eq:compmagtensor}
  m_{xx,i} & = & \left( p_i\mu_i \right) \left( 1 - \phi_{yy,i} \right), 
  \nonumber \\ 
  m_{yy,i} & = & \left( p_i\mu_i \right) \left( 1 - \phi_{xx,i} \right), 
  \nonumber \\ 
  m_{xx,i} & = & \left( p_i\mu_i \right) \phi_{xy,i}.
\end{eqnarray}
While the expression for the magnification $\mu_i$ is given in
equation~\eqref{eq:fluxeq}, those for the partial derivatives of the
deflection potential follow as
\begin{eqnarray}
  \label{eq:partderivpot}
  \phi_{xx,i} & = &  \left\{ \right. 
  \left[ \alpha^2 \cos\theta_i^2 - \alpha (\cos\theta_i^2 - \sin\theta_i^2) \right] F_i
  \nonumber \\ 
  & & - 2 (\alpha-1) \sin\theta_i \cos\theta_i F'_i 
  + \sin\theta_i^2 F''_i 
  \left. \right\}  R_i^{\alpha-2}, 
  \nonumber \\
  \phi_{yy,i} & = &  \left\{ \right. 
  \left[ \alpha^2 \sin\theta_i^2 + \alpha (\cos\theta_i^2 - \sin\theta_i^2) \right] F_i
  \nonumber \\ 
  & & + 2 (\alpha-1) \sin\theta_i \cos\theta_i F'_i 
  + \cos\theta_i^2 F''_i 
  \left. \right\}  R_i^{\alpha-2}, 
  \nonumber \\ 
  \phi_{xy,i} & = &  \left\{ \right. 
  \alpha (\alpha-2) \sin\theta_i \cos\theta_i  F_i
  - \sin\theta_i \cos\theta_i  F''_i 
  \nonumber \\ 
  & & + (\alpha-1) (\cos\theta_i^2 - \sin\theta_i^2) F'_i 
  \left. \right\}  R_i^{\alpha-2}.
\end{eqnarray}
Next, we turn the predicted magnifications $\tilde{\mu}_i$ through
equation~\eqref{eq:fluxeq} per image into predictions for the flux
ratios $\tilde{f}_j \equiv \tilde{\mu}_1/\tilde{\mu}_{1+j}$. We
compare them to the observed flux ratios $f_j$ with corresponding
errors $\Delta f_j$ by evaluating
\begin{equation}
  \label{eq:Chi2flux}
  \chi_\mathrm{flx}^2 = 
  \sum_{j=1}^3 \frac{(f_j-\tilde{f_j})^2}{\Delta f_j^2}.
\end{equation}
We can now find the best-fit parameters by minimizing
\begin{equation}
  \label{eq:chi2tot}
  \chi^2 = \chi_\mathrm{src}^2 + \chi_\mathrm{flx}^2 
  + \lambda \sum_{m \ge 3}^n (c_m^2+s_m^2),
\end{equation}
with $\lambda \ge 0$ a constant. The last term guides the solution
towards a smooth, realistic lens model by penalizing (strong)
deviations from an elliptic shape.

For a range of slopes $0.5<\alpha<1.5$, the fits of the above lens
model with $n=4$ Fourier terms are shown in Fig.~\ref{fig:ecfit}.
Whereas the predictions of the source position remain within the
observed errors over the full range of slopes, the recovery of the
observed radio fluxes places constraints on the slope. Dividing the
corresponding $\chi^2$ values by the ten degrees of freedom (eleven
constraints minus the slope parameter) provides a quality of fit,
shown in the lower-right panel of Fig.~\ref{fig:ecfit}. The minimum
value $\chi^2/10 \simeq 1$ shows that for $\alpha \simeq 1.0$ a good
overall best-fit model is obtained. The corresponding
$99$\,\%-confidence interval (dotted line) is $0.9 \lesssim \alpha
\lesssim 1.1$.

With $n<4$ Fourier terms the fits are significantly worse ($\chi^2/10
> 2.5$), and with $n=5$ Fourier terms the fit does not improve (for
$n>5$ the fitting problem becomes underdetermined). If we replace the
radio fluxes by the optical fluxes from the \castles\ website, no
acceptable lens model fit is possible ($\chi^2/10 > 30$).  On the
other hand, the mid-infrared fluxes from \cite{2000ApJ...545..657A}
yield similar good fits as the radio fluxes. The best-fit slope
$\alpha=0.9$ is lower because the fluxes of image A and C are
respectively about 10\,\% and 30\,\% lower than in the radio.  Image C
is the faintest in both mid-infrared and radio, while image D is the
faintest in the optical. Since the radio is expected to be (much) less
affected by differential extinction and microlensing, we adopt the
corresponding lens models.

In the investigation below, we use the lens models with slopes
$\alpha=\{0.8,0.9,1.0,1.1,1.2\}$, shown in Fig.~\ref{fig:ecmod}, and
with best-fit parameters given in Table~\ref{tab:ecmod}. The critical
curves $\Rc(\theta)$ are obtained by solving the
(quadratic) equation~\eqref{eq:fluxeq} for infinite magnification,
i.e., for vanishing left-hand-side. The caustics then follow upon
substitution of $\Rc(\theta)$ in the lens
equation~\eqref{eq:lenseq}. We see in Fig.~\ref{fig:ecmod} that even
the two models outside the $99$\,\%-confidence interval for $\alpha$
are relatively smooth as expected for a realistic galaxy model.

The surface mass density of the lens models follows from Poisson's
equation as
\begin{equation}
\label{eq:surfdens}
    \Sigma(R,\theta) = \Sigc \,
    \frac12 \left[ \alpha^2 F(\theta) + F''(\theta) \right] R^{\alpha-2},
\end{equation}
with the critical surface mass density defined as
\begin{equation}
  \label{eq:critsurfdens}
    \Sigc = \frac{c^2\,D_s}{4\pi G\,D_l\,D_{ls}},
\end{equation}
where $c$ is the speed of light and $D_l$, $D_s$ and $D_{ls}$ are the
(angular diameter) distance to the lens galaxy, the quasar source and
the distance from lens to source, respectively. 

The (projected) mass within the critical curve
\begin{equation}
  \label{eq:masscrit}
  \Mc = 
  \Sigc
  \int_0^{2\pi} \frac{1}{2\alpha} 
  \left[ \alpha^2 F(\theta) + F''(\theta) \right]
  \Rc(\theta)^\alpha \du\,\theta,
\end{equation}
is in general close to the projected mass within the Einstein radius.
The latter is the radius $\Rein$ for which the projected mass is equal
to the Einstein mass defined as $\Mein = \Sigc \pi \Rein^2$. Because
$\Rein$ is independent of $\theta$, all higher order Fourier terms
average out, and we are left with $\Rein^{2-\alpha} = \alpha c_0/2$,
so that the Einstein mass becomes
\begin{equation}
  \label{eq:masseinstein}
  \Mein = \Sigc \, \pi \left( \frac{\alpha c_0}{2} \right)^{2/(2-\alpha)},
\end{equation}
Since we express all (projected) coordinates on the plane of the sky
in arcseconds, including the radius $R$, we multiply
expressions~\eqref{eq:masscrit} and~\eqref{eq:masseinstein} by $(D_l
\, \pi/0.648)^2$ to convert from arcseconds to pc for a given (angular
diameter) distance $D_l$ to the lens galaxy in Mpc.

From the last two columns in Table~\ref{tab:ecmod}, we see indeed that
$\Mc \approx \Mein = 1.54 \times 10^{10}$\,\Msun, nearly independent
of the slope $\alpha$. Taking into account an inverse scaling with the
Hubble constant of $H_0 = 73$\,\kmsM, our values are within a few per
cent of previous measurements
\citep[e.g.][]{1992AJ....104..959R,1994AJ....108.1156W,
  1998ApJ...495..609C, 1998MNRAS.295..488S, 2002MNRAS.334..621T}.

\section{Results}
\label{sec:results}


\subsection{Luminous versus total mass distribution}
\label{sec:massdistr}

\begin{table}
\begin{center}
\caption{Multi-Gaussian Expansion parameters\label{tab:mgepar}}
\begin{tabular}{c*{3}{r}c*{3}{r}}
\hline
\hline
    & \multicolumn{3}{c}{$I$-band surface brightness} &
    & \multicolumn{3}{c}{$\alpha=1.0$ lens model} \\
\cline{2-4} \cline{6-8}
$i$ & $\log\mathrm{I}_0$ & $\log\sigma'$ & $q'$ &
    & $\log\Sigma_{0}$    & $\log\sigma'$ & $q'$ \\
\hline
 1 & 4.329 & -1.564 & 0.700 & & 5.114 & -1.398 & 0.645\\
 2 & 3.935 & -0.942 & 0.700 & & 4.556 & -1.037 & 0.660\\
 3 & 3.606 & -0.627 & 0.700 & & 4.238 & -0.779 & 0.662\\
 4 & 3.293 & -0.239 & 0.700 & & 3.928 & -0.544 & 0.678\\
 5 & 3.005 & -0.043 & 0.700 & & 3.628 & -0.350 & 0.673\\
 6 & 2.845 &  0.230 & 0.700 & & 3.459 & -0.186 & 0.675\\
 7 & 2.261 &  0.587 & 0.700 & & 3.354 & -0.001 & 0.675\\
 8 & 2.160 &  0.917 & 0.414 & & 3.208 &  0.220 & 0.675\\
 9 & 1.334 &  1.135 & 0.700 & & 3.069 &  0.496 & 0.675\\
10 & -     &  -     & -     & & 2.991 &  1.000 & 0.675\\                       
\hline
\end{tabular}
\tablecomments{The parameters of the Gaussians in the MGE fit to the
  \hst/WFPC2/F814W $I$-band image of the surface brightness (columns
  2--4, ``deconvolved'' with the WFPC2 PSF), and to the surface mass
  density of the overall best-fit lens model with slope $\alpha = 1.0$
  (columns 5--7) of the lens galaxy in the Einstein Cross (see also
  Fig.~\ref{fig:mgesurf}).  Columns two and five give for each
  Gaussian component respectively the central surface mass density (in
  \Msunpcsq) and the central surface brightness (in \Lsunpcsq),
  columns three and six the dispersion (in arcsec) along the major
  axis, and columns four and seven the observed flattening.}
\end{center}
\end{table}

In the left and middle panel of Fig.~\ref{fig:mgesurf} we show MGE
fits (contours) to respectively the $V$-band and the $I$-band \hst\ 
image, obtained with the software of \cite{2002MNRAS.333..400C}, while
masking the quasar images. For each fit we assume the same position
angle (PA) for the Gaussians, consistent with the adopted axisymmetry
in the dynamical models.  Although the bar and spirals cannot be
reproduced, it provides a good description of the disk in the outer
parts (left panel) and reproduces well the bulge in the inner parts
(middle panel). Similarly, we derive MGE fits to the surface mass
density in equation~\eqref{eq:surfdens} of the best-fit lens models
with slopes $\alpha$ from 0.8 to 1.2. In the right panel of
Fig.~\ref{fig:mgesurf}, we show the MGE fit (solid contours) to the
surface mass density (dashed contours) of the overall best-fit lens
model with slope $\alpha = 1.0$. The corresponding parameters are
given in Table~\ref{tab:mgepar} (columns 5--7), together with the
parameters of the MGE fit to the $I$-band surface brightness (columns
2--4). The latter parameters are ``deconvolved'' using a MGE model
fitted to the WFPC2 PSF as determined with Tiny Tim \citep{TinyTim}.

The Gaussians in the MGE fit to the $I$-band surface brightness have a
PA of $\sim 70$\dgr, which is similar to the PA of $\simeq 67$\dgr\ in
the MGE fit to the surface mass density of lens models. Both these
values are consistent with the measurements by
\cite{1988AJ.....95.1331Y}, who found a PA of $\simeq 67$\dgr\ for the
axis through quasar images C and D, bracketed by a PA of $\simeq
77$\dgr\ for the outer disk and a PA of $\simeq 39$\dgr\ for the bar
\citep[see also Fig.~1 of][]{2002MNRAS.334..621T}.

\begin{figure*}
  \begin{center}
    \includegraphics[width=1.0\textwidth]{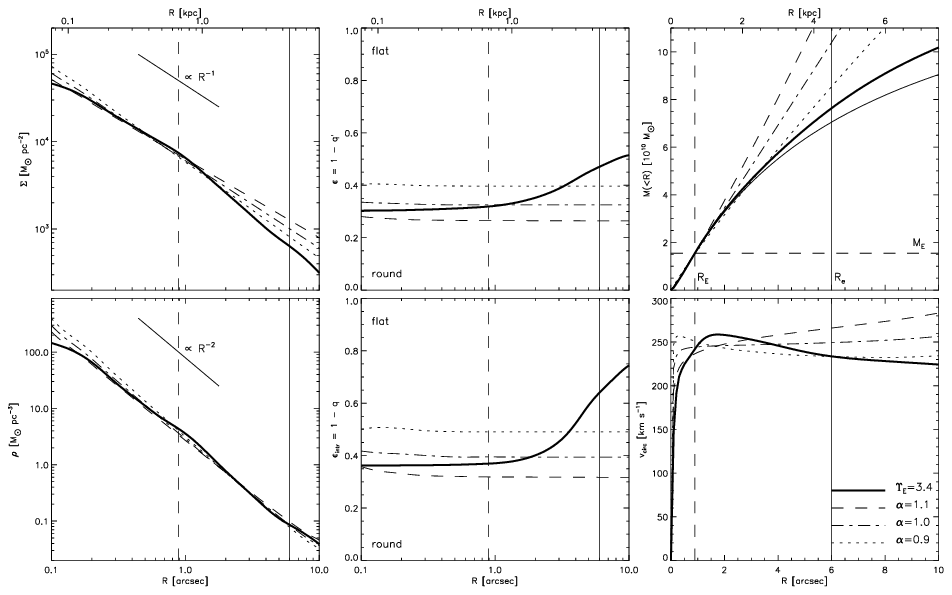}
  \end{center}
  \caption{ Radial profiles of the mass distribution of the lens
    galaxy in the Einstein Cross as function of radius $R$, in arcsec
    at the bottom and in kpc in at the top. In each panel the dashed
    and solid vertical lines indicate respectively the Einstein radius
    $\Rein=0.89$\arcsec\ and the effective radius $R_e=6$\arcsec.  The
    \emph{top panels} shows from left to right the surface mass
    density profile $\Sigma$, the corresponding projected ellipticity
    $\epsilon = 1 - q'$, and the mass $M$ enclosed within the
    projected radius $R$ along the major-axis on the plane of the sky.
    The dotted, dash-dotted and dashed curves are from the best-fit
    lens models with (fixed) slopes $\alpha$ of respectively $0.9$,
    $1.0$ and $1.1$ (see \S~\ref{sec:lensmodel}).  The thick solid
    curves are from the observed ($I$-band) surface brightness,
    multiplied with a constant mass-to-light ratio $\MLein =
    3.4$\,\MLsunI.  The latter is obtained from dividing the Einstein
    mass (indicated by the horizontal dashed line in the top-right
    panel) by the (projected) luminosity within the Einstein
    radius. The thin solid curve in the top-right panel shows the
    enclosed projected mass of a de Vaucouleurs profile fitted to the
    bulge which is dominating the mass in the inner parts of the lens
    galaxy. The \emph{bottom panels} shows from left to right the mass
    density profile $\rho$, the corresponding intrinsic ellipticity
    $\epsilon_\mathrm{intr} = 1 - q$, and the circular velocity
    $v_\mathrm{circ}$ as function of the radius $R$ in the meridional
    plane. These intrinsic quantities are obtained from the MGE fits
    to the projected density, under the assumption of oblate
    axisymmetry and for an inclination $i=68$\dgr\ (see also
    Appendix~\ref{sec:appmge}).}
  \label{fig:ecmassdistr}
\end{figure*}

As shown in Fig.~\ref{fig:ecmassdistr}, not only the orientation, but
also the shape of the luminous density distribution as inferred from
the surface brightness, is very similar to that of the total mass
density distribution as derived from the lens models. The left and
middle panel in the top row show respectively the profile and the
projected ellipticity $\epsilon = 1 - q'$ of the surface mass density
as function of the major axis on the sky-plane. The thin curves are
for the lens models with fixed slope $\alpha$ (as indicated in the
bottom-right panel). The thick solid curve is for the $I$-band surface
brightness, multiplied with a constant mass-to-light ratio $\MLein =
3.4$\,\MLsunI. The latter is derived from dividing the projected mass
from the lens models (shown in the top-right panel of
Fig.~\ref{fig:ecmassdistr}) within the Einstein radius, i.e., the
Einstein mass $\Mein$, by the projected luminosity within the Einstein
radius $\Rein$.

We see that around $\Rein$ (indicated by the dashed vertical line),
where the lens models are best constrained by the observed quasar
image positions and relative flux ratios, both the slope and the
ellipticity of the \emph{projected} luminous and total mass density
are nearly the same.  The same holds true for the corresponding
\emph{intrinsic} mass densities, of which the profiles and
ellipticities $\epsilon_\mathrm{intr} = 1 - q$ are shown in the left
and middle panel in the bottom row of Fig.~\ref{fig:ecmassdistr}.
These (analytic) deprojections (see Appendix~\ref{sec:appmge}) are
under the assumption of oblate axisymmetry and for a given inclination
which we derive from the flattening of the disk in the outer parts. We
find from the MGE fit to the $V$-band surface brightness (left panel
of Fig.~\ref{fig:mgesurf}), that Gaussian components with a measured
flattening as small as $q' \simeq 0.4$ are required for an acceptable
fit. This sets a lower limit to the inclination of $i \gtrsim 66$\dgr
\citep[significantly larger than $i \simeq 60$\dgr\ found
by][]{1989AJ.....98.1989I}. Adopting a lower limit for the
\textit{intrinsic} flattening of the disk of $q = 0.15$
\citep[e.g.][]{1992MNRAS.258..404L}, we then obtain an inclination of
$i \simeq 68$\dgr.

Around $\Rein$ the luminosity density is close to isothermal ($\rho
\propto R^{-2}$ intrinsic or $\Sigma \propto R^{-1}$ in projection)
like the mass density of the overall best-fit lens model. Towards the
center the slope of the luminosity density becomes shallower. The
(fixed) slope of the lens model is not anymore (well) constrained by
the quasar images towards the center, neither for radii a few times
$\Rein$.  At these larger radii also the luminosity density is not only
due to the bulge, but the bar and disk start contributing, as can also
be seen from the increase in the ellipticity. Nevertheless, within a
radius $R \lesssim 4$\arcsec, the intrinsic luminous and total mass
distribution are very similar, i.e., mass follows light. This means we
can infer the gravitational potential either from the observed surface
brightness adopting a (constant) total mass-to-light ratio $\MLtot$,
or directly from the best-fit lens model (see also
Appendix~\ref{sec:appmge}). The corresponding circular velocity
curves, defined as $v_\mathrm{circ}^2 = R \partial\Phi/\partial R$ in the
equatorial plane, are shown in the bottom-right panel of
Fig.~\ref{fig:ecmassdistr}, using $\MLein = 3.4$\,\MLsunI\ as above.

\subsection{Axisymmetric Jeans models}
\label{sec:axijeansmodels}

\begin{figure*}
  \begin{center}
    \includegraphics[width=1.0\textwidth]{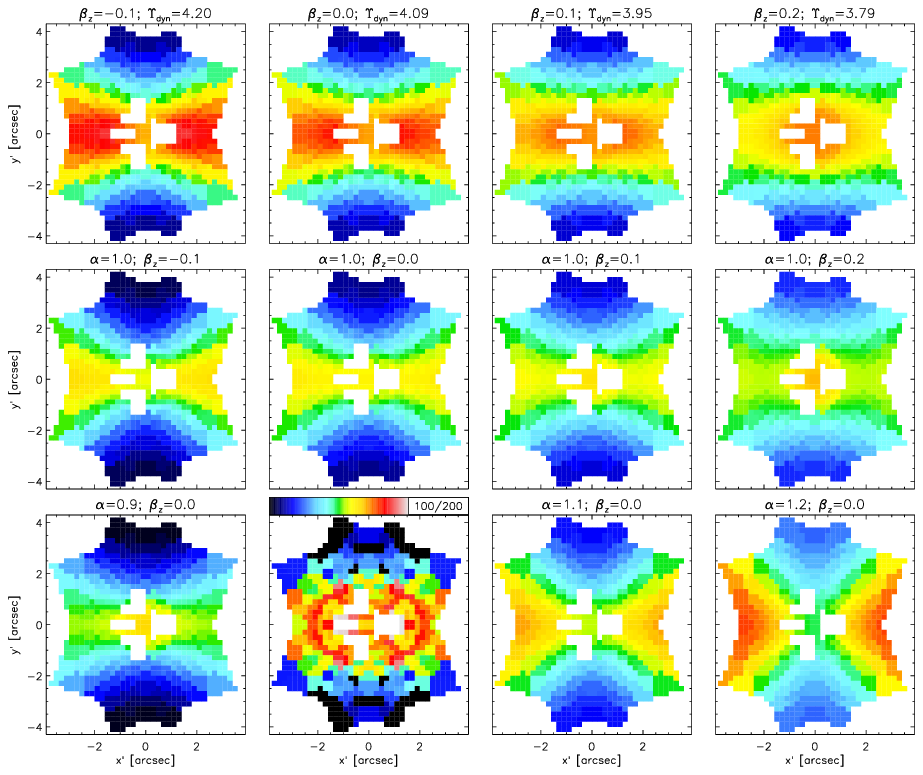}
  \end{center}
  \caption{Axisymmetric Jeans model of the lens galaxy in the Einstein
    Cross.  The second panel in the bottom row shows the square root
    of the combination $V^2+\sigma^2$ of the observed mean
    line-of-sight velocity $V$ and dispersion $\sigma$ maps
    (Fig.~\ref{fig:eckinmaps}). The (linear) scale in \kms\ is
    indicated by the color bar and limits at the top, and is the same
    for all other panels, which show the line-of-sight second-order
    velocity moment as predicted by the axisymmetric Jeans models
    (\S~\ref{sec:dynamicalmodel} and
    Appendix~\ref{sec:appmgeaxijeans}). In the \emph{top row}, the
    gravitational potential is inferred from the $I$-band surface
    brightness, assuming that mass follows light. The corresponding
    (constant) dynamical mass-to-light ratio $\MLdyn$ that provides
    the best-fit to the observations is indicated at the top, together
    with the assumed value for the anisotropy parameter, $\beta_z = 1
    - \overline{v_z^2}/\overline{v_R^2}$, increasing from left to
    right. The latter parameter represents the (constant) flattening
    of the velocity ellipsoid in the meridional plane.  The
    \emph{middle row} follows the same sequence of increasing $\beta_Z$
    values, but the gravitational potential is now computed directly,
    i.e., without mass-to-light ratio conversion, from the overall
    best-fit lens model with slope $\alpha = 1.0$.  In the
    \emph{bottom row}, besides the observations in the second panel,
    the predictions for different slopes $\alpha$ are shown, with
    $\beta_Z=0$, corresponding to an isotropic velocity distribution in
    the meridional plane.}
  \label{fig:jeansmodel}
\end{figure*}

In Fig.~\ref{fig:jeansmodel}, the second panel in the bottom row shows
the square root of the combination $V^2+\sigma^2$ of the observed (and
symmetrized) mean line-of-sight velocity $V$ and dispersion $\sigma$
maps of the lens galaxy (Fig.~\ref{fig:eckinmaps}, two panels on the
right). 
The other panels in Fig.~\ref{fig:jeansmodel} show the line-of-sight
second-order velocity moment as predicted by the JAM models
\citep[][]{2008MNRAS.390...71C}, which are based on the solution of
the axisymmetric Jeans equations as summarized in
Appendix~\ref{sec:appmge}.
With a central velocity dispersion of $\sigma_0 \simeq 170$\,\kms, the
$\Mbh-\sigma$ relation as given by \cite{2002ApJ...574..740T} predicts
a central black hole of mass $\Mbh \simeq 7.0 \times 10^7$\,\Msun\ for
the lens galaxy.  Since the corresponding sphere-of-influence
$\Mbh/\sigma^2 \simeq 0.01$\arcsec\ is not resolved by our kinematic
data, we cannot fit for it, but do include the predicted BH in the
Jeans models (as an additional Gaussian). Finally, we convolve the
predicted second-order velocity moment with the Gaussian PSF of the
kinematic data with FWHM$=1$\arcsec. The (linear) scale in \kms, as
indicated by the color bar and limits at the top of the panel with the
observations, is the same in all panels of Fig.~\ref{fig:jeansmodel}.

The axisymmetric Jeans models in the top row use the gravitational
potential inferred from the $I$-band surface brightness, assuming that
mass follows light. The corresponding (constant) dynamical
mass-to-light ratio $\MLdyn$ that provides the best-fit to the
observations is indicated at the top, together with the assumed value
for the anisotropy parameter $\beta_Z$, increasing from left to right.
The corresponding increase in the (constant) flattening of the
velocity ellipsoid in the meridional plane (or decrease in
$\overline{v_z^2}/\overline{v_R^2}$), has two main effects: a less
pinched ``butterfly'' shape and a weaker gradient parallel to the
(vertically aligned) short axis. Comparing with the observations, we
see that at lower $\beta_Z$ values the pinching is too strong, while at
higher $\beta_z$ the gradient along the short axis is too shallow.
Therefore, it is not unexpected that the ``best-fit'' is when $\beta_Z
\simeq 0$ and $\MLdyn \simeq 4.0$\,\MLsunI\ (in the $I$-band), but
with reduced $\chi^2 \simeq 3$ significantly above unity.

Part of the latter mismatch might be due to (systematic) discrepancies
in the data, in particular due to the challenging measurement of the
velocity dispersion. The average uncertainty in the observed
second-order velocity moment is about $10$\,\kms, which translates
into an error of around $12$ per cent in $\MLdyn$, or about
$0.5$\,\MLsunI. However, the uncertainties in the observed
second-order velocity moment vary quite a lot, from only $\sim
2$\,\kms\ in the center to $\sim 20$\,\kms\ a the edge of the \gmos\ 
field. And although we were very careful in disentangling the
contribution from the quasar and excluded the regions at the quasar
images, the measured velocity dispersion around these regions might
still be systematically affected by (broadening due to) residual light
from the quasar source. This possibly explains the observed excess in
the observed second-order velocity moment around the quasar images
with respect to the accurately measured central value. At the same
time, such an excess might mimic a steeper gradient in the observed
second-order velocity moment. Indeed, in all Jeans models in the top
row of Fig.~\ref{fig:jeansmodel} the prediction at the center is
significantly higher than the observed value. As a result, we expect
that the dynamical mass-to-light ratio is overestimated.

If, alternatively, we infer the gravitational potential directly,
i.e., without the need for a mass-to-light ratio, from the
(deprojected) surface mass density of the best-fit lens models, the
axisymmetric Jeans models predict maps of the second-order velocity
moments that are similar in shape but with lower values near the
center. This is shown in the middle row of Fig.~\ref{fig:jeansmodel}
for the overall best-fit lens model with slope $\alpha=1$, and for the
same range in anisotropy parameters $\beta_Z$. The formal best-fit is
obtained for $\beta_z \simeq 0$, but the observed central value is
more closely matched in case of mild velocity anisotropy in the
meridional plane with $\beta_Z \gtrsim 0.1$.  For $\alpha=0.9$. i.e.,
density slopes steeper then isothermal, the observed central value is
best matched when $\beta_Z \simeq 0$, with at the same time a steeper
gradient along the short-axis, as shown in the first panel in the
bottom row of Fig.~\ref{fig:jeansmodel}. The last two panels show that
for $\alpha>1$, velocity isotropy in the meridional plane ($\beta_z =
0$) results in a central value much lower than the observed
second-order velocity moment.

We conclude that for the lens models with slopes $\alpha = 1.0 \pm
0.1$, as constrained by the lensing geometry, and for a mild
anisotropic velocity distribution in the meridional plane with
$\beta_Z = 0.1 \pm 0.1$, the predictions of the second-order velocity
moments match the accurately measured central value, and follow the
gradient within the observational uncertainties outside the region
that might be affected by (residual) quasar light. The latter results
in broadening of the velocity dispersion, so that the best-fit
dynamical mass-to-light ratios we derive when we infer the
gravitational potential from the surface brightness, are likely
overestimated.  Fitting instead only the unaffected and accurately
measured second-order velocity moment in the center, we find for
$\beta_Z = 0.1$ a best-fit value of $\MLdyn \simeq 3.7$\,\MLsunI, with
an error around $0.5$\,\MLsunI. On the other hand, both the Einstein
mass and luminosity within the Einstein radius are well constrained,
resulting in at most a few per cent error in $\MLein = 3.4$\,\MLsunI.
Therefore, in what follows, we adopt the latter as the total
mass-to-light ratio in the inner bulge-dominated region of the lens
galaxy.

In the above axisymmetric Jeans models we fix the inclination to $i =
68$\dgr\ determined from the observed ellipticity of the outer disk
(see \S~\ref{sec:massdistr}). When we vary the inclination over a
reasonable range above the lower limit $i \gtrsim 66$\dgr, there is no
significant change in the above results, which is expected because the
inclination only has a weak effect on the mass-to-light ratio
\citep[see also Fig.~4 of][]{2006MNRAS.366.1126C}.

\subsection{Stellar mass-to-light ratio}
\label{sec:stellarml}

We have shown that in the bulge-dominated inner region of the lens
galaxy, the total mass distribution closely follows the light
distribution. This does not mean there is no dark matter present,
since if dark matter follows (nearly) the same distribution it might
contribute to part of the above estimated total mass-to-light ratio
$\MLtot \simeq 3.4$\,\MLsunI. To determine this possible
\emph{constant} dark matter fraction, we need to measure the stellar
mass-to-light ratio $\MLstar$. To this end we use the single stellar
population (SSP) models of \cite{1996ApJS..106..307V}, adopting as
initial reference the \cite{2001MNRAS.322..231K} initial mass function
(IMF) with lower mass cut-off $0.01$\,\Msun (faintest star
$0.09$\,\Msun) and upper mass cut-off of $120$\,\Msun.

We derive a high-S/N spectrum from our data cube by collapsing the
central spectra that are unaffected by the quasar images. Even so, the
Ca~II triplet is equally well fitted by SSP models of nearly all ages
and metallicities of about $\feh = \pm 0.2$ around solar, which leaves
the stellar mass-to-light ratio nearly unconstrained.  Therefore, we
also derived colors from archive HST images (PI: Kochanek). 
After matching all images, using both the quasar positions and
isophotes of the lens galaxy, we compute the magnitudes in a circular
aperture of radius $R = 0.2$\arcsec\ (well within the quasar images at
$R \simeq 0.9$\arcsec). 
%
%
%
After correcting for a Galactic extinction of $E(B-V)=0.071$
\citep{1998ApJ...500..525S}, we derive in the F555W, F675W, F814W
(WFPC2), F160W and F205W (NICMOS2) filters magnitudes of 16.83, 16.03,
15.53, 13.80, and 13.51, respectively. We convert the latter to
\emph{rest frame} magnitudes of 16.72, 16.13, 15.50, 13.70, 13.59 in
Johnson $V$, $R$, $I$, $H$ and $K$ filters, respectively. Instead of
the standard K-correction, we use the method described in \S~3 of
\cite{2003MNRAS.344..924V} to take simultaneously into account the
filter responses and the redshift of the lens galaxy.

The resulting colors are best fitted by a SSP model with age $t \simeq
8$\,Gyr and metallicity $\feh \simeq 0.2$, corresponding to an
$I$-band stellar mass-to-light ratio $\MLstar \simeq 3.3$\,\MLsunI.
The $1\sigma$ confidence limits yield a range in $t$ from 7 to
14\,Gyr, and in $\feh$ from $0.0$ to $0.3$, corresponding to a range
in $\MLstar$ from $2.8$ to $4.1$\,\MLsunI. The lower limit implies at
most $\sim 20$ per cent of dark matter within $R \la 4$\arcsec, but
the results are fully consistent with no dark matter at all. The SSP
models of \cite{1998MNRAS.300..872M, 2005MNRAS.362..799M} with a
Kroupa IMF yield similar results. Adopting a
\cite{1955ApJ...121..161S} IMF with the same lower and upper mass
cut-offs, increases $\MLstar$ by about $30$ per cent. This not only
implies no dark matter, but even for the lower limit $\MLstar >
\MLtot$, which is unphysical. This is consistent with evidences for
both late-type galaxies \citep{2001ApJ...550..212B} and early-type
galaxies \citep{2006MNRAS.366.1126C} that, if the IMF is universal, a
Salpeter IMF is excluded, whereas a Kroupa IMF matches the
observations \citep[see][for a review]{2007iuse.book..107D}.

\section{Discussion and conclusions}
\label{sec:discconcl}

We used the \gmos-North integral-field spectrograph to obtain
two-dimensional stellar kinematics of the lens galaxy in the Einstein
Cross. In addition to the four bright quasar images and the distance
of the lens galaxy ($D_l = 155$\,Mpc), in particular the presence of
sky lines in the observed Ca~II triplet region made the extraction of
the absorption line kinematics challenging. Even so, we were able to
derive high-quality line-of-sight velocity $V$ and dispersion $\sigma$
maps of the bulge-dominated inner region $R \lesssim 4$\arcsec,
reaching about two-thirds of the effective radius $\Reff \simeq
6$\arcsec\ of this early-type spiral galaxy. The $V$ map shows regular
rotation up to $\sim 100$\,\kms\ around the minor axis of the bulge,
consistent with axisymmetry. The $\sigma$ map shows a weak gradient
increasing towards a central ($R<1$\arcsec) value of $170 \pm
9$\,\kms.

The only other direct (single) measurement of the velocity dispersion
is $215 \pm 30$\,\kms\ by \cite{1992ApJ...386L..43F}. Adopting their
central aperture of 0.7\arcsec$\times$0.4\arcsec\ at a position angle
of 39\dgr\ (along the bar), our extracted spectrum is shown in
Fig.~\ref{fig:censpec}. The lower panel shows sky-subtracted galaxy
spectrum (black curve) and our best-fit composite stellar population
model (blue curve), yielding a velocity dispersion in this aperture of
172\,\kms. The other model (red curve) has the velocity dispersion
fixed to 215\,\kms\ as measured by \cite{1992ApJ...386L..43F}. Since
the observing conditions were similar with seeing around $0.5$\arcsec,
we can directly compare both measurements. We see from residuals at
the bottom and the quoted reduced $\chi^2$-values that 172\,\kms
provides a better fit than 215\,\kms, particularly in fitting the
strongest of the three absorption lines.

In addition, for a singular isothermal sphere lens model, we can use
the relation $\Delta\theta=8\pi(\sigma_\mathrm{SIS}/c)^2D_{ls}/D_s$
\citep[e.g.][]{2000ApJ...543..131K} with a separation $\Delta\theta =
1.8$\arcsec\ of the four quasar images, to obtain a simple estimate
for the dispersion of $\sigma_\mathrm{SIS} \simeq 180$\,\kms. 
Adopting a spherical Jeans model with a range in velocity anisotropy,
\cite{2003MNRAS.344..924V} convert this to a central velocity
dispersion of $168 \pm 17$\,\kms\ within a circular ``Coma'' aperture
(with a diameter of 3.4\arcsec at the distance of the Coma
cluster). At the (angular diameter) distance $D_l = 155$\,Mpc of the
lens galaxy, this corresponds to a circular aperture with radius of
1\arcsec, so that it can be compared directly to our central value of
$170 \pm 9$\,\kms.
Finally, King and de Vaucouleurs models of \cite{1988AJ.....96.1570K}
predict a similar value of $\sim 166$\,\kms, and also
\cite{1999MNRAS.309..641B} find a value of $165 \pm 23$\,\kms\ based
on their two \HI\ rotation curve measurements.
Hence, it is likely that the long-slit measurement by
\cite{1992ApJ...386L..43F} is affected by the bright quasar images,
whereas our spatially resolved measurements allow for a clean(er)
separation of the quasar contribution.

A large variety of different lens models have been constructed for the
Einstein Cross, most of which fit the positions of the quasar images
but not their relative flux ratios. Adopting the scale-free lens model
of \cite{2003MNRAS.345.1351E}, we found that fitting at the same time
also the (radio) flux ratios constrained the slope of the total mass
surface density $\Sigma \propto R^{\alpha-2}$ to be $\alpha = 1.0 \pm
0.1$. The total mass within the Einstein radius $\Rein = 0.89$\arcsec,
i.e., the Einstein mass, is $\Mein = 1.54 \times 10^{10}$\,\Msun,
nearly independent of the slope $\alpha$, and consistent with previous
measurements in the literature. Dividing by the projected luminosity
within $\Rein$, as measured from the observed $I$-band surface
brightness, we obtained a mass-to-light ratio of $\MLein =
3.4$\,\MLsunI, with an error of at most a few per cent.

We determined the mass-to-light ratio in an additional, independent
way by fitting dynamical models to the (combined) observed $V$ and
$\sigma$ maps. We used the solution of the axisymmetric Jeans
equations to predict this second-order velocity moment, with the
gravitational potential inferred from the deprojected $I$-band surface
brightness, assuming that mass follows light. We expect the resulting
best-fit constant mass-to-light ratio to be an overestimation due to
possible residual contribution from the quasar light. Even so,
restricting the fit to the unaffected and accurately measured central
second-order velocity moment, we found $\MLdyn = 3.7 \pm
0.5$\,\MLsunI. When we used instead the gravitational potential that
follows directly, i.e., without mass-to-light ratio conversion, from
the surface mass density of the best-fit lens models, we arrived at a
similar prediction of the second-order velocity moment. We showed that
the reason is that the luminous and total mass distribution, as
inferred from respectively the surface brightness and lens model, are
very similar. This implies that the inner region of the lens galaxy
mass (closely) follows light, with a total mass-to-light ratio $\MLtot
\simeq 3.4$\,\MLsunI.

By fitting single stellar population models to measured colors of the
center of the lens galaxy, we estimated an $I$-band stellar
mass-to-light ratio $\MLstar$ from 2.8 to 4.1\,\MLsunI. Although a
\emph{constant} dark matter fraction of 20 per cent is thus not
excluded, it is likely that dark matter does not play a significant
role in the inner region of this early-type spiral galaxy of
luminosity $\sim L\star$. This is consistent with indications that
less-luminous early-type galaxies \citep[e.g.][]{2001AJ....121.1936G,
  2003ApJ...595...29R, 2006MNRAS.366.1126C, 2007MNRAS.382..657T,
  2008ApJ...684..248B} and late-type galaxies
\citep[e.g.][]{1996MNRAS.281...27P, 2000AJ....120.2884P,
  2001ApJ...550..212B, 2006ApJ...643..804K, 2009MNRAS.400.1665W} are
dominated by the stellar mass inside their central regions. This is
different from dwarf galaxies as well as in (at least the outer parts
of) giant elliptical and spiral galaxies where dark matter is expected
to be ubiquitous.

The constraint $\alpha = 1.0 \pm 0.1$ on the slope of the lens model
implies that the intrinsic total mass density is close to isothermal,
consistent with previous studies of lens galaxies
\citep[e.g.][]{2006ApJ...649..599K, 2009ApJ...703L..51K}. However, one
has to be very careful not to over-interpret this result, not only due
to assumptions in the modeling, but most of all because the slope is
only (well) constrained around the Einstein radius $\Rein$. If indeed
mass follows light in the inner region of this lens galaxy, the
(deprojected) surface brightness indicates deviations from isothermal
towards the center where the slope becomes shallower as well as a
possible steepening at larger radii. Interestingly, the construction
of realistic galaxy density profiles with a stellar and dark matter
component that are both non-isothermal, shows that the combined slope
and corresponding lensing properties are nevertheless consistent with
isothermal around $\Rein$ \citep{2009MNRAS.398..607V}.  With more
extended images, e.g. in the case of galaxy-galaxy lensing, one can
place a stronger constraint on the total mass distribution of the lens
galaxy \citep[e.g.][]{2007ApJ...666..726B, 2009ApJ...691..277S}. Since
the extension in these cases is still mostly tangential, constraining
a large radial range is in particular possible when lensing occurs in
multiple (redshift) planes, resulting in images at different Einstein
radii \citep[e.g.][]{2008ApJ...677.1046G}.

Nevertheless, already the quasar images provide an accurate constraint
on the total mass within $\Rein$, nearly independent of the details of
the lens model \citep[e.g.][]{1991ApJ...373..354K,
  2001MNRAS.327.1260E}.  Therefore, at least around this radius no
higher-order velocity moments are needed to break the mass-anisotropy
degeneracy. Towards the center the surface brightness increases
steeply, so that high enough S/N to measure velocity moments beyond
$V$ and $\sigma$ might be achievable even at higher redshift. In the
outer parts, the degeneracy might be (partially) broken by combining
the kinematics of stars and/or discrete tracers, such as globular
clusters and planetary nebulae, with total mass estimates from hot
X-ray gas and/or weak lensing measurement
\citep[e.g.][]{2007ApJ...667..176G}. In the current and even more in
the upcoming extensive and deep photometric surveys, numerous (strong)
gravitational lensing systems will be discovered.  They provide
important and independent constraints on the total mass distribution
in galaxies, especially in combination with (resolved) stellar
kinematics, as we showed in this paper \citep[see
also][]{2007ApJ...666..726B, 2008MNRAS.384..987C,
  2009MNRAS.399...21B}.


\acknowledgments

We are grateful to Jean-Ren\'e Roy and Matt Mountain for granting us
director's discretionary time for this project and for generous
hospitality in Hilo to TdZ. We thank Ed Turner for initial support of
this project and Tracy Beck for efficient and cheerful assistance.
We thank the referee for constructive comments on this work.
GvdV acknowledges support provided by NASA through Hubble Fellowship
grant HST-HF-01202.01-A awarded by the Space Telescope Science
Institute, which is operated by the Association of Universities for
Research in Astronomy, Inc., for NASA, under contract NAS 5-26555.  MC
acknowledges support from a STFC Advanced Fellowship (PP/D005574/1).
Based on observations obtained at the Gemini Observatory, which is
operated by the Association of Universities for Research in Astronomy,
Inc., under a cooperative agreement with the NSF on behalf of the
Gemini partnership: the National Science Foundation (United States),
the Particle Physics and Astronomy Research Council (United Kingdom),
the National Research Council (Canada), CONICYT (Chile), the
Australian Research Council (Australia), CNPq (Brazil), and CONICET
(Argentina).


\appendix

\section{Multi-Gaussian Expansion}
\label{sec:appmge}

We summarize the (numerically) convenient expressions of the
axisymmetric intrinsic density and gravitational potential in the case
of a Multi-Gaussian Expansion (MGE) of the surface density
\citep{1992A&A...253..366M, 1994A&A...285..723E}. Next, we show that
also the gravitational lensing properties can be readily computed, so
that the MGE method can be used to efficiently construct general lens
models. Finally, we give the line-of-sight second-order velocity
moment as a solution of the axisymmetric Jeans equations which is
discussed in detail in \cite{2008MNRAS.390...71C}.

\subsection{Axisymmetric density and potential}
\label{sec:appmgedenspot}

We parameterize the surface brightness $I(x',y')$ by a sum of
Gaussian components
\begin{equation}
  \label{eq:mgeIxyproj}
  I_j(x',y') = I_{0,j}
  \exp\left\{ -\frac{1}{2{\sigma'_j}^2} 
    \left[ x'^2 + \frac{y'^2}{{q'_j}^2} \right] \right\}.
\end{equation}
each with three parameters: the central surface brightness $I_{0,j}$,
the dispersion $\sigma'_j$ along the major $x'$-axis and the
flattening $q'_j$. In case of an (oblate) axisymmetric system viewed
at an inclination $i>0$, the corresponding intrinsic luminosity
density is
\begin{equation}
  \label{eq:mgenuRz}
  \nu_j(R,z) = 
  \frac{q'_j I_{0,j}}{\sqrt{2\pi}\sigma_j q_j} 
  \exp\left\{ -\frac{1}{2\sigma_j^2} 
    \left[ R^2 + \frac{z^2}{q_j^2} \right] \right\},
\end{equation}
with intrinsic dispersion $\sigma_j = \sigma'_j$ and intrinsic
flattening $q_j$ given by $q_j^2 \sin^2i =  {q'_j}^2 - \cos^2i$.

The intrinsic mass density follows as $\rho_j = \MLj \, \nu_j$, where
the mass-to-light ratio $\MLj$ per Gaussian component is a free
parameter. Alternatively, if the surface mass density $\Sigma(x',y')$
is available (e.g.\ from gravitational lensing), an
MGE-parameterization as in \eqref{eq:mgeIxyproj} directly yields
$\rho_j$ as in \eqref{eq:mgenuRz}, but with $I_{0,j}$ in both
expressions replaced by the central surface mass density
$\Sigma_{0,j}$.

The corresponding gravitational potential follows upon (numerical)
evaluation of \citep[][eq.~39]{1994A&A...285..723E}
\begin{equation}
  \label{eq:mgepotRz}
  \Phi_j(R,z) =
  -\frac{2 G M_j}{\sqrt{2\pi}\sigma_j} \int_0^1 
  \mathcal{F}_j(u) \, \du u,
\end{equation}
where we have introduced
\begin{equation}
  \label{eq:mgedefFj}
  \mathcal{F}_j(u) = \exp\left\{ -\frac{u^2}{2\sigma_j^2}  
    \left[ R^2 + \frac{z^2}{{\mathcal{Q}_j^2(u)}} \right] 
  \right\} \frac{1}{\mathcal{Q}_j(u)},
\end{equation}
and $\mathcal{Q}_j^2(u) = 1-(1-q_j^2) \, u^2$.
The total mass per Gaussian component is given by $M_j = 2 \pi
{\sigma'_j}^2 {q'_j} \Sigma_{0,j}$, with $\Sigma_{0,j} = \MLj \,
I_{0,j}$ when the mass density is inferred from the surface
brightness.

In the latter case, we have implicitly assumed a \emph{constant}
mass-to-light ratio $\MLj$ per Gaussian component by taking it outside
the integral. Nevertheless, we can still mimic a (radially) varying
mass-to-light ratio by considering the $\MLj$ of the Gaussian
components as free parameters \citep[see e.g.][]{2006A&A...445..513V,
  2006ApJ...641..852V}.  However, it is common in dynamical studies of
the inner parts of galaxies \citep[e.g.][]{2006MNRAS.366.1126C} to
assume the total mass-to-light ratio to be constant, i.e., $\MLj =
\MLtot$ for each Gaussian component $j$. Since $\MLtot$ may be larger
than the stellar mass-to-light ratio $\MLstar$, this still allows for
possible dark matter contribution, but with a constant fraction.

\subsection{Axisymmetric lens model}
\label{sec:appmgeaxilens}

Under the thin-lens approximation, the gravitational lensing
properties of a galaxy are characterized by the deflection potential
$\phi(x',y')$ and its (partial) derivatives. The deflection potential
follows from projecting the potential along the line-of-sight or by
solving the two-dimensional Poisson equation $\nabla^2 \phi = 2
\kappa$. Here, $\kappa = \Sigma/\Sigc$ is the normalized surface mass
density with the critical lensing value $\Sigma_c$ defined in
equation~\eqref{eq:critsurfdens}.

Similar to the surface brightness in equation~\eqref{eq:mgeIxyproj},
we parameterize this so-called ``convergence'' $\kappa(x',y')$ by a
sum of Gaussian components
\begin{equation}
  \label{eq:mge_kappa_k}
  \kappa_k(x',y') = \frac{\Sigma_{0,j}}{\Sigc}
  \exp\left\{ -\frac{1}{2{{\sigma'}_k}^2} 
    \left[ {x'}^2 + \frac{{y'}^2}{{q'_k}^2} \right] \right\}.
\end{equation}
The corresponding deflection potential is then
\begin{equation}
  \label{eq:mge_phi_k}
  \phi_k(x',y') = - \frac{M_k}{\pi\Sigc} 
  \int_0^1 {\mathcal{F}'_k}(u) \, \frac{\du u}{u},
\end{equation}
where $M_k = 2 \pi {\sigma'_k}^2 {q'_k} \Sigma_{0,k}$ is the total
mass per Gaussian component, and we have introduced
\begin{equation}
  \label{eq:mge_Fp_k}
  {\mathcal{F}'_k}(u) = \exp\left\{ -\frac{u^2}{2{\sigma'}_k^2}  
    \left[ {x'}^2 + \frac{{y'}^2}{{\mathcal{Q}'_k}^2(u)} \right]
  \right\} \frac{1}{{\mathcal{Q}'_k}(u)},
\end{equation}
and ${\mathcal{Q}'_k}^2(u) = 1-(1-{q'_k}^2) \, u^2$.

The first-order partial derivatives of $\phi_k(x',y')$ follow as
\begin{eqnarray}
  \frac{\partial \phi_k}{\partial {x'}} 
  & = & 
  {x'} \frac{M_k}{\pi{\sigma'_k}^2\Sigc}
  \int_0^1 {\mathcal{F}'_k}(u) \, 
  u \, \du u,
  \nonumber \\ \label{eq:mge_dphi_k}
  \frac{\partial \phi_k}{\partial {y'}} 
  & = & 
  {y'} \frac{M_k}{\pi{\sigma'_k}^2\Sigc}
  \int_0^1 {\mathcal{F}'_k}(u) \,
  \frac{u \, \du u}{{\mathcal{Q}'_k}^2(u)},
\end{eqnarray}
whereas the second-order partial derivatives are given by
\begin{eqnarray}
  \frac{\partial^2 \phi_k}{\partial {x'}^2} 
  & = & 
  \frac{M_k}{\pi{\sigma'_k}^2\Sigc}
  \int_0^1 {\mathcal{F}'_k}(u)
  \left[ 1 - \frac{{x'}^2}{{\sigma'_k}^2} \, u^2 \right]
  u \, \du u,
  \nonumber \\
  \frac{\partial^2 \phi_k}{\partial {y'}^2} 
  & = &
  \frac{M_k}{\pi{\sigma'_k}^2\Sigc}
  \int_0^1 {\mathcal{F}'_k}(u)
  \left[ 1 - \frac{{x'}^2}{{\sigma'_k}^2}
  \frac{u^2}{{\mathcal{Q}'_k}^2(u)} \right]
  \frac{u \, \du u}{{\mathcal{Q}'_k}^2(u)},
  \nonumber \\ \label{eq:mge_dphi2_k}
  \frac{\partial^2 \phi_k}{\partial {x'} \partial {y'}} 
  & = & 
  \frac{M_k}{\pi{\sigma'_k}^2\Sigc}
  \int_0^1 {\mathcal{F}'_k}(u)
  \left[ \frac{{x'}{y'}}{{\sigma'_k}^2} \, u^2 \right]
  \frac{u \, \du u}{{\mathcal{Q}'_k}^2(u)}.
\end{eqnarray}
The above single integrals can be readily evaluated numerically, so
that the lensing properties follow in a straightforward way.
The image positions $(x',y')$ are related to the source position
$(\xi,\eta)$ by the lens equation
\begin{equation}
  \label{eq:mge_lenseq}
  \xi = x' - \sum_k \frac{\partial \phi_k}{\partial {x'}},
  \qquad
  \eta = y' - \sum_k \frac{\partial \phi_k}{\partial {y'}}.
\end{equation}
Given the parity $p$ of the image, the magnification $\mu$ is given by
\begin{equation}
  \label{eq:mge_magnification}
  \frac{1}{p\mu} = 
    \left(1 - \sum_k \frac{\partial^2 \phi_k}{\partial {x'}^2} \right) 
    \left(1 - \sum_k \frac{\partial^2 \phi_k}{\partial {y'}^2} \right) 
    - \left( \sum_k \frac{\partial^2 \phi_k}
      {\partial {x'} \partial {y'}} \right).
\end{equation}
The Einstein radius $\Rein$ follows from solving $\bar{\kappa}(\Rein) =
1$, where $\bar{\kappa}(R')$ is the average convergence within the
projected radius $R'$. After substituting $x'=R'\cos\theta'$ and
$y'=R'\sin\theta'$ in equation~\eqref{eq:mge_kappa_k} and performing the
integral over $R'$, we are left with
\begin{equation}
  \label{eq:mge_avkappa}
  \bar{\kappa}_k(R') = 
  \frac{{\sigma'_k}^2\Sigma_{0,k}}{\pi {R'}^2 \Sigc}
  \int_0^{2\pi} \left\{ 
    1 - \exp\left[-\frac{{R'}^2{\mathcal{P}'}(\theta')}{2{\sigma'_k}^2}\right]
    \right\} \frac{\du\theta'}{{\mathcal{P}'}(\theta')},
\end{equation}
where ${\mathcal{P}'}(\theta') = \cos^2\theta' +
\sin^2\theta'/{q'_k}^2$. The mass within the Einstein radius then
follows as $\Mein = \Sigc \pi \Rein^2$.

\subsection{Axisymmetric Jeans equations}
\label{sec:appmgeaxijeans}

We consider a luminous component with Gaussian intrinsic luminosity
density $\nu_j$ given by \eqref{eq:mgenuRz}, of which the observed
kinematics trace the underlying gravitational potential $\Phi = \sum_k
\Phi_k$, with the sum over all (luminous and dark) components $\Phi_k$
given by \eqref{eq:mgepotRz}. 

Assuming that the velocity ellipsoid is aligned with the cylindrical
$(R,\phi,z)$ coordinate system, so that $\overline{v_R v_z}=0$, we can
readily solve Jeans equation~\eqref{eq:cylcbevz} as
\citep[][eq.~42]{1994A&A...285..723E}
\begin{equation}
  \label{eq:cylvz2mge}
  [\overline{v_z^2}]_j =
  4 \pi G
  \sum_k \frac{q'_k \Sigma_{0,k}}{\sqrt{2\pi} \sigma_k} 
  \int_0^1 q_j^2 \sigma_j^2
  \frac{\mathcal{F}_k(u) u^2}{1-F_{jk}u^2} \; \du u,
\end{equation}
with $\mathcal{F}_k(u)$ defined in \eqref{eq:mgedefFj} and
\begin{equation}
  \label{eq:cyldefFjku}
  F_{jk} = 1 - q_k^2 - q_j^2 \sigma_j^2/\sigma_k^2.
\end{equation}
If we next also assume a constant flattening of the velocity ellipsoid
in the meridional plane, we can write $[\overline{v_R^2}]_j = b_j \,
[\overline{v_z^2}]_j$, and solve Jeans equation~\eqref{eq:cylcbevR} as
\begin{equation}
  \label{eq:cylvphi2mge}
  [\overline{v_\phi^2}]_j =
  4 \pi G
  \sum_k \frac{q'_k \Sigma_{0,k}}{\sqrt{2\pi} \sigma_k}
  \int_0^1 \left[ b_j q_j^2 \sigma_j^2 + \mathcal{G}_{jk}(u) R^2 \right]
  \frac{\mathcal{F}_k(u) u^2}{1-F_{jk}u^2} \; \du u,
\end{equation}
where we have introduced
\begin{equation}
  \label{eq:cyldefGjku}
  \mathcal{G}_{jk}(u) =
  1 - b_jq_j^2 - [b_j(1-q_k^2) + (1-b_j)F_{jk}]u^2.
\end{equation}
Since the constant $b_j$ may be different for each luminous component
$j$, the total anisotropy in the meridional plane
\begin{equation}
  \label{eq:cylbetamge}
  \beta_Z = 1 - \frac{\overline{v_z^2}}{\overline{v_R^2}}
  = 1 - \frac{\sum_j \nu_j [\overline{v_z^2}]_j}
  {\sum_j b_j \nu_j [\overline{v_z^2}]_j},
\end{equation}
is allowed to vary throughout the system. In case of isotropy in the
meridional plane all $b_j = 1$ and thus $\beta_Z = 0$, corresponding
to a two-integral DF $f(E,L_z)$ \citep[][]{1994A&A...285..723E}.

After substitution of these expressions in
equation~\eqref{eq:cylvlosmge}, the line-of-sight integral can be
solved, resulting in \citep[][eq.~28]{2008MNRAS.390...71C}
\begin{multline}
  \label{eq:cylvlos2mge}
  [\overline{v_\mathrm{los}^2}]_j = 
  \frac{4 \pi^{3/2} G}{I_j(x',y')} \,
  \frac{q'_j I_{0,j}}{\sqrt{2\pi}\sigma_j q_j}
  \sum_k \frac{q'_k \Sigma_{0,k}}{\sqrt{2\pi} \sigma_k}
  \int_0^1 \frac{q_j^2 \sigma_j^2 (\cos^2i + b_j\sin^2i) +
    \mathcal{G}_{jk}(u) x'^2\sin^2i}
  {(1-F_{jk}u^2)\sqrt{[1-(1-q_k^2)u^2]
      (\mathcal{A}+\mathcal{B}\cos^2i)}}
  \\ \times
  u^2 \exp\left\{ - \mathcal{A} \left[ x'^2 + 
      \frac{(\mathcal{A}+\mathcal{B})y'^2}
      {\mathcal{A}+\mathcal{B}\cos^2i}
    \right] \right\} \, \du u,
\end{multline}
where $\mathcal{A}$ and $\mathcal{B}$ are functions of $u$ defined as
\begin{eqnarray}
  \label{eq:cyldefA}
  \mathcal{A} & = & \frac12 \left( 
    \frac{1}{\sigma_j^2} + \frac{u^2}{\sigma_k^2} \right),
  \\
  \label{eq:cyldefB}
  \mathcal{B} & = & \frac12 \left( 
    \frac{1-q_j^2}{q_j^2\sigma_j^2} + 
    \frac{(1-q_k^2)u^4}{[1-(1-q_k^2)u^2]\sigma_k^2} \right).
\end{eqnarray}
The remaining single integral can be readily evaluated numerically.

This provides a direct prediction of the combination $V^2 + \sigma^2$
of the observed mean line-of-sight velocity $V$ and dispersion
$\sigma$ at a given position $(x',y')$ on the sky-plane, through the
(luminosity weighted) sum $\overline{v_\mathrm{los}^2}(x',y') = \sum_j
I_j [\overline{v_\mathrm{los}^2}]_j / \sum_j I_j$ over the
corresponding luminous components.




\end{document}